\documentclass[a4paper,11pt]{article}
\pdfoutput=1 

\usepackage{jheppub} 

\usepackage[T1]{fontenc} 
\usepackage{amsmath}
\usepackage{mathrsfs}
\usepackage{comment}
\usepackage{color}
\usepackage{subcaption}
\usepackage{physics}

\makeatletter
\g@addto@macro\bfseries{\boldmath}
\makeatother

\newcommand{\wt}[1]{\widetilde{#1}}
\newcommand{\msf}[1]{\mathsf{#1}}

\newcommand{\mc}[1]{\mathcal{#1}}

\newcommand{\tl}{\tilde}

\renewcommand{\i}{\text{i}}
\newcommand{\e}{\text{e}}
\newcommand{\sgn}{\text{sign}}
\newcommand{\sign}{\text{sign}}
\newcommand{\ud}{\text{d}}

\newcommand{\cG}{\mathcal{G}}

\newcommand{\mq}{{q}}
\newcommand{\vol}{\text{vol}}

\newcommand{\hvar}{\delta}

\title{\boldmath Emergent Time in Hamiltonian General Relativity}

\author{Anurag Kaushal,}
\author{Naveen S.~Prabhakar, and}
\author{Spenta R.~Wadia}

\affiliation{International Centre for Theoretical Sciences-Tata Institute of Fundamental Research, Shivakote, Bengaluru 560089, India.}

\emailAdd{anuragkaushal314@gmail.com}
\emailAdd{naveen.s.prabhakar@gmail.com}
\emailAdd{spenta.wadia@icts.res.in}

\abstract{In this paper we introduce a definition of time that emerges
  in terms of the geometry of the configuration space of a dynamical
  system. We illustrate this, using the Hamilton-Jacobi equation, in
  various examples: particle mechanics on a fixed energy surface;
  non-Abelian gauge theories for compact semi-simple Lie groups where
  the Gauss law presents new features; and General Relativity in $d+1$
  dimensions with $d$ the dimension of space. The discussion in
  General Relativity is like the non-abelian gauge theory case except
  for the indefiniteness of the de Witt metric in the
  Einstein-Hamilton-Jacobi equation, which we discuss in some
  detail. We illustrate the general formula for the emergent time in
  various examples including de Sitter spacetime and asymptotically
  AdS spacetimes. }

\begin{document} 
\maketitle
\flushbottom

\section{Introduction}

The invention of the idea of time and its measurement is a fundamental
ingredient in the description of dynamical systems. Newtonian
mechanics describes the motion of a point particle in a three
dimensional space in terms of coordinates which are functions of a
universal parameter $t$ that is measurable by a system (like a clock)
which also obeys the laws of motion. The final state of a system is
determined by the laws of motion given a set of initial
conditions. Time here is universal in the sense that all clocks can be
simultaneously synchronized, a fact that is modified by special
relativity due to the constancy of the speed of light for all inertial
observers and time and space are related by linear Lorentz
transformations. Relativistic field theory retains the notion of
specifying initial data on a constant time slice and then evolving it
via the field equations.

General Relativity (GR) radically changes the notion of time because
the theory is invariant under spacetime diffeomorphisms. There is no
obvious choice of a fixed time slice in spacetime and evolution of
initial data using Einstein's equations. To address this problem Dirac
introduced the theory of constrained Hamiltonian systems
\cite{Dirac:1950pj,dirac2001lectures}. Dirac \cite{Dirac:1958sc} and
Arnowitt-Deser-Misner (ADM) \cite{Arnowitt:1962hi}, gave a description
of the time foliation of space-time in terms of a 3-geometry embedded
in $3+1$ dim spacetime. The Hamiltonian $H$ of GR is a linear
combination of first class constraints $\mc{H}_\perp(x)$,
$\mc{H}^i(x)$: $H = \int\ud^3x (N \mc{H}_\perp + N_i \mc{H}^i)$, where
$N$ and $N_i$ are the lapse function and the shift vector
respectively. The lapse and shift are usually fixed by an appropriate
choice of gauge, i.e., a choice of coordinates on the spacetime. The
semi-classical quantum theory is then described by the Wheeler-de Witt
equation $H|\Psi\rangle = 0$ (perhaps more correctly to be called the
Schrodinger-Wheeler-de Witt equation).

One of the aims of this investigation is to give an intrinsic
definition of `time' that emerges from the geometry of the
configuration space of metrics and matter fields that makes no appeal
to the existence of an external time. Our method is based on the
Einstein-Hamilton-Jacobi equation \cite{Peres1962OnCP}, that also
follows from the Wheeler-de Witt equation $H|\Psi \rangle = 0$ in the
semiclassical limit.

Before we develop the aforementioned notion of time for GR, we
illustrate the main idea for particle mechanics and non-abelian gauge
theories in Minkowski spacetime. The analog of the Wheeler-de Witt
equation is the time-independent Schrodinger equation
$H |\Psi\rangle = E|\Psi\rangle$. The classical dynamics is thus on a
fixed energy surface in phase space, and is described by the time
independent Hamilton-Jacobi equation -- which itself arises in the
semiclassical limit of the Schrodinger equation. In this case, there
is no \emph{a priori} notion of time since the system is on a constant
energy surface. We will show that time emerges in terms of the
positive path length of the Riemannian geometry of the configuration
space. By virtue of its definition in terms of the geometry of
configuration space, this notion of time can also be applied to
dynamics that is classically forbidden but whose trajectories exist as
imaginary time instantons in configuration space \cite{Wadia:1979yu}.

Taking over these ideas to GR in $d+1$ dimensions ($d \geq 2$), one
encounters the difficulty posed by the fact that the metric in the
configuration space of $d$-metrics that follows from
Einstein-Hamilton-Jacobi equation -- called the de Witt metric -- has
indefinite signature. This implies that paths in configuration space
can be spacelike, timelike or null with respect to the de Witt
metric. We find that the notion of time that follows from the
Hamilton-Jacobi equation can be defined for paths that are either
spacelike or timelike, and the same method cannot be applied for null
paths in configuration space. The notion of time thus derived from a
study of the Einstein-Hamilton-Jacobi equation is given by the simple
$d$-diffeomorphism invariant formula
\begin{equation}\label{unitau}
  \ud \tau = \frac{\ud s_\epsilon}{2\sqrt{-\epsilon V}}\ ,
\end{equation}
where $\ud s_\epsilon$ is the infinitesimal positive line element in
the configuration space of metrics on the spatial slice and
$\epsilon = +1,-1,0$ depending on whether the above path is spacelike,
timelike or null with respect to the de Witt metric on the
configuration space. The quantity $V$ is the `potential' function for
general relativity given by
$V = \int\ud^dx\, N \sqrt{g} (R - 2\Lambda) + \text{matter
  contributions}$, where $N$ is the lapse function, $g_{ij}$ is the
metric on a spatial slice on which the integral above is carried out,
and $\Lambda$ is the cosmological constant. In particle mechanics and
gauge theories, there is an analogous formula for time along classical
paths where the denominator is replaced by $\sqrt{2(E-V)}$ where $E$
is the energy of the configuration and $V$ is the potential
function(al). The proof that \eqref{unitau} is the `correct'
definition of time lies in the fact that we reproduce corresponding
familiar equations of motion in each of the situations (particle
mechanics, non-abelian gauge theories and general relativity) above.

We illustrate the formula \eqref{unitau} in the context of de Sitter
spacetime in Section \ref{examples}, where it becomes a simple
function of the volume of the spatial slices. We also discuss the case
of asymptotically Anti de Sitter spacetimes in Section
\ref{adsexample} where the ability to always choose zero mean
curvature foliations ensure that the path in configuration space is
spacelike, and hence serves as a concrete illustration of our
procedure in deriving \eqref{unitau}.

The idea of time in general relativity has been debated upon in
various contexts. There are many proposals which apply in restricted
situations, like the (log of) the volume of spatial slices in
homogeneous cosmologies \cite{Misner:1969hg,
  Misner:1969ae,Chakraborty:2023yed}, the mean extrinsic curvature of
spatial slices for spacetimes that allow constant mean curvature
slicings \cite{York:1972sj}, an external time based on past volume of
a spatial slice \cite{Sorkin:1987cd, Unruh:1988in, Unruh:1989db,
  Henneaux:1989zc}, the proper time of dust worldlines in the case of
general relativity coupled to dust \cite{Brown:1994py}, and so on. The
notion of time in quantum gravity has also been explored by studying
the Wheeler-de Witt equation \cite{DeWitt:1967yk, Lapchinsky:1979fd,
  Banks:1984cw, Halliwell:1984eu, Vilenkin:1988yd}; there are also
ideas that suggest that quantum gravity is timeless
\cite{Hartle:1983ai, Hawking:1983hj, Page:1983uc, Hawking:1985bk,
  Hartle:1988zs, Halliwell:1990qr}. See the reviews
\cite{Kuchar:1991qf, Isham:1992ms, Kiefer:2021zdq, Maniccia:2022iqa},
and references therein for a comprehensive discussion.

More recently, in the context of AdS/CFT, a notion of time that is
appropriate for infalling observers in black hole spacetimes has been
proposed in \cite{Leutheusser:2021frk} based on the algebra of
operators in the dual CFT. It has been shown in
\cite{Araujo-Regado:2022gvw} that a certain irrelevant $T\bar{T}$-like
deformation of the dual CFT gives rise to an emergent notion of time
in asymptotically AdS spacetimes. In the context of two dimensional de
Sitter JT gravity, the dilaton field has been used as a clock to study
solutions of the Wheeler-de Witt equation \cite{Nanda:2023wne}.

In the Polyakov formulation of string theory where the two dimensional
world sheet is described by a unitary conformal field theory coupled
to a dynamical metric, the Liouville mode emerges as a time or space
dimension in target space, depending on the value of the central
charge of the conformal field theory \cite{Das:1988ds, Das:1989da}. In
particular it was shown in \cite{Dhar:1989km} that 25 massless scalars
coupled to two dimensional gravity exactly reproduces the Veneziano
amplitude in $25+1$ dimensions, giving a direct evidence of the
emergence of time from dynamical two dimensional gravity in world
sheet string theory. Dynamic processes like tachyon condensation can
be used to give a notion of time in string theory, with the tachyon
field treated as the time variable \cite{Sen:2002qa}.

\section{Particle Mechanics}\label{particle}

Consider a dynamical system with phase space coordinates $q$, $p$,
with Hamiltonian $H(q,p)$. The time-independent Hamilton-Jacobi
equation for Hamilton's characteristic function $W(q)$ is
\begin{equation}\label{timeindHJ}
	H\left(q, \frac{\ud W}{\ud q}\right) = E\ .
\end{equation}
This arises as the semi-classical $\hbar \to 0$ limit of the
time-independent Schrodinger equation. Let us illustrate this for a
particle in a one dimensional potential $V(q)$ with Hamiltonian
\begin{equation}\label{Hparticle}
	H(q,p) = \frac{p^2}{2m} + V(q)\ .
\end{equation}
The time-independent Schrodinger equation for the wavefunction
$\psi(q)$ is
\begin{equation}\label{schparticle}
	-\frac{\hbar^2}{2 m }\frac{\ud^2 \psi(q)}{\ud q^2} + V(q) \psi(q) = E \psi(q)\ .
\end{equation}
Substituting $\psi(q) = \e^{\i W(q) / \hbar} \chi(q)$, we get at
leading order in $\hbar$, viz., $\hbar^0$,
\begin{equation}
	\frac{1}{2m}\left(\frac{\ud W}{\ud q}\right)^2 + V(q) = E\ ,
\end{equation}
which is \eqref{timeindHJ} with the expression for $H$ in
\eqref{Hparticle}.
We focus on the problem of a particle on an $n$-dimensional Riemannian
manifold $X$ with coordinates $q^i$, $i=1,\ldots,n$, and metric
\begin{equation}\label{tanmet}
  \ud s^2 = \sum_{i,j=1}^n g_{ij}(q) \ud q^i \ud q^j\ .
\end{equation}
Let the corresponding conjugate momenta be $p_i$.\footnote{We suppress
  the index $i$ on $q^i$ and $p_i$ frequently to avoid clutter in
  notation. We also follow the Einstein summation convention
  throughout the paper where repeated indices are assumed to be summed
  over unless otherwise indicated, and seldom display explicit
  summation symbols.} The Hamiltonian is
\begin{equation}\label{ppHam}
	H(q,p) = \frac{1}{2}\sum_{i,j=1}^n g^{ij}(q) p_i p_j + V(q)\ ,
\end{equation}
where $g^{ij}$ is the inverse of the metric $g_{ij}$. The
Hamilton-Jacobi equation then becomes
\begin{equation}\label{hamjacgen}
	\frac{1}{2}  g^{ij}(q)  \frac{\partial W}{\partial q^i} \frac{\partial W}{\partial q^j} + V(q) = E\ .
\end{equation}
\textbf{Note:} It is no accident that the inverse metric appears in
the Hamilton-Jacobi equation \eqref{hamjacgen}: since
$\partial W / \partial q^i$ can be thought of as a cotangent vector,
the natural metric on cotangent vectors is $g^{ij}$.

The Hamilton-Jacobi equation can be solved for $W(q)$ as follows. Let
$W[q(s)]$ be a functional on the set of \emph{all} paths $q(s)$ in
configuration space which is the Riemannian manifold $X$. The path
parameter $s$ is the distance along the path measured with respect to
the metric \eqref{tanmet} on $X$. Suppose $\msf{q}(s)$ is an extremum
of $W[q(s)]$ with end-point $q^i$. Consider the functional
$W[\msf{q}(s)]$ evaluated on the extremum path: this is automatically
a function of the end-point of the extremum path, and \emph{this
  function is designated as $W(q)$}.

The expression for the functional $W[\msf{q}(s)]$ is deduced by
demanding that $W(q)$ -- the functional evaluated on an extremum path
treated as a function of the end-point of the path -- satisfy the
Hamilton-Jacobi equation \eqref{hamjacgen}. We can choose coordinates
in a neighbourhood of the path where one of the coordinate axes is
along the tangent vector $v^i = \ud \msf{q}^i / \ud s$ and the other $n-1$
axes $n_A^i$, $A = 2,\ldots,n$, are orthogonal to it with respect to
$g_{ij}$. Since $\msf{q}(s)$ is an extremal path, the directional
derivatives of $W$ along the normal axes are zero because the
functional $W$ is stationary along such deformations:
\begin{equation}
  n_A^i \frac{\partial W}{\partial q^i} = 0\ .
\end{equation}
The only non-zero directional derivative is along the tangent vector
to the path. Thus, on the extremal path in configuration space, one
can write
\begin{equation}\label{Wtangder}
  \frac{\partial W}{\partial q^i} = g_{ij}\big(\msf{q}(s_1)\big) \frac{\ud \msf{q}^j(s_1)}{\ud s} \frac{\ud W}{\ud s_1}\ ,
\end{equation}
where $s_1$ is the value of the path parameter $s$ at the end-point of
the extremal path, i.e., $\msf{q}(s_1) = q^i$. It is easy to see from
\eqref{tanmet} that the tangent vector $v^i = \ud \msf{q}^i / \ud s$ has
unit norm with respect to the metric $g_{ij}$. Plugging in
\eqref{Wtangder} into the Hamilton-Jacobi equation \eqref{hamjacgen},
we get
\begin{equation}\label{hamjacgen1}
  \frac{1}{2}  \left( \frac{\ud W}{\ud s_1}\right)^2 + V(q) = E\ .
\end{equation}
We can thus write
\begin{equation}\label{Wsol}
	W(q) = W[\msf{q}(s)] = \int_{s_0}^{s_1} \ud s \sqrt{2\big(E- V(\msf{q}(s))\big)}\ ,
\end{equation}
where $\msf{q}(s)$ is the extremal path. Since the above quantity can
be evaluated for any given path in configuration space, we extend the
definition of the functional $W[q(s)]$ to be the right hand side above
for all paths in configuration space.

We now recover the equation for the extremal path by setting
$\delta W = 0$ under path variations
$\mq(s) \to \mq(s) + \delta \mq(s)$ which are zero at the beginning
and end points. This procedure can be found in Landau and Lifshitz,
Volume I: Classical Mechanics \cite[\S 44, Eq.(44.10) and the
associated Problem]{landau1976mechanics}\footnote{The action
  \eqref{Wsol} is referred to as the abbreviated action
  $S_0 = \int p \ud q = \int \sqrt{2(E-V) g_{ij} \ud \mq^i \ud \mq^j}$
  in \cite[Eq.(44.9)]{landau1976mechanics}. See also
  \cite{Banks:1973ps,Banks:1973uca,Gervais:1977nv}.}. Recall that
$\ud s^2 = g_{ij}(\mq(s)) \ud \mq^i \ud \mq^j$. We then have
\begin{equation}\label{qmsvar}
	\delta \ud s = \frac{1}{2}\ud s \left(g_{ij,k} \frac{\ud \mq^i}{\ud s} \frac{\ud \mq^j}{\ud s} \delta \mq^k + 2g_{ij} \frac{\ud \mq^i}{\ud s} \frac{\ud \delta \mq^j}{\ud s} \right)\ ,
\end{equation}
where $g_{ij,k} = \partial g_{ij} / \partial q^k$, and
\begin{equation}\label{qmVvar}
	\delta \sqrt{2(E-V)} = -\frac{1}{\sqrt{2(E-V)}} \frac{\partial V}{\partial \mq^i} \delta \mq^i\ ,
\end{equation}
so that
\begin{align}
  \delta W
  &=\int_{s_0}^{s_1} \left(\delta \ud s \sqrt{2(E-V)} + \ud s\, \delta\sqrt{2(E-V)}\right)\ ,\nonumber\\
  &= \int_{s_0}^{s_1}   \ud s\left(\frac{1}{2} g_{ij,k} \frac{\ud \mq^i}{\ud s} \frac{\ud \mq^j}{\ud s} \delta \mq^k + g_{ij} \frac{\ud \mq^i}{\ud s} \frac{\ud \delta \mq^j}{\ud s}\right)\sqrt{2(E-V)}-\int_{s_0}^{s_1} \ud s \frac{1}{\sqrt{2(E-V)}} \frac{\partial V}{\partial \mq^i} \delta \mq^i\ .
\end{align}
Now, looking at the various occurrences of the factor $\sqrt{2(E-V)}$,
we define a new parameter $\tau$ such that
\begin{equation}\label{staudef}
  \boxed{    \frac{\ud}{\ud\tau} =  \sqrt{2(E-V)} \frac{\ud}{\ud s}\ ,\quad \ud\tau = \frac{\ud s}{\sqrt{2(E-V)}}\ .}
\end{equation}
The variation of $W$ then takes the form
\begin{align}\label{Wvartau}
  \delta W &= \int_{\tau_0}^{\tau_1}   \ud \tau\left(\frac{1}{2} g_{ij,k} \frac{\ud \mq^i}{\ud \tau} \frac{\ud  \mq^j}{\ud \tau} \delta \mq^k + g_{ij} \frac{\ud \mq^i}{\ud \tau} \frac{\ud \delta \mq^j}{\ud \tau}- \frac{\partial V}{\partial \mq^i} \delta \mq^i\right)\ ,\nonumber\\
           &= \int_{\tau_0}^{\tau_1}   \ud \tau\left(\frac{1}{2} g_{ij,k} \frac{\ud \mq^i}{\ud \tau} \frac{\ud \mq^j}{\ud \tau} \delta \mq^k -\frac{\ud}{\ud\tau} \left(g_{ik} \frac{\ud \mq^i}{\ud \tau}\right)  - \frac{\partial V}{\partial \mq^k}\right) \delta \mq^k + g_{ik} \frac{\ud q^i}{\ud \tau}\delta q^k\bigg|_{\tau_0}^{\tau_1}\ ,
\end{align}
where we have integrated by parts the $\ud / \ud \tau$ in the second
step and written the total derivative as a boundary term. The boundary
term above is zero since $\delta q = 0$ at the beginning and end
points of the path. The equation for the path obtained by setting
$\delta W = 0$ is
\begin{equation}
	\frac{1}{2}  g_{ij,k}  \frac{\ud \mq^i}{\ud \tau}\frac{\ud \mq^j}{\ud \tau} -    \frac{\ud}{\ud \tau} \left( g_{ik}\frac{\ud \mq^i}{\ud \tau}\right) - \frac{\partial V}{\partial \mq^k} = 0\ .
\end{equation}
Pushing in the derivative in the second term and contracting with
$g^{\ell k}$, we get the geodesic equation in a potential
\begin{equation}\label{eom}
	\frac{\ud^2 \mq^i}{\ud \tau^2} + \Gamma^{i}_{jk} \frac{\ud \mq^j}{\ud \tau}\frac{\ud \mq^k}{\ud \tau} + g^{ij}\frac{\partial V}{\partial \mq^j} = 0\ ,
\end{equation}
where
$\Gamma^i_{jk} = \frac{1}{2} g^{i \ell} ( g_{j\ell,k} + g_{\ell k,j} -
g_{jk,\ell})$ is the Christoffel connection. Thus, we get the usual
Euler-Lagrange equation of motion for a particle, but now with the
`time' $\tau$ \eqref{staudef} being defined in terms of the
configuration space variables and the potential. From the dynamics of
the particle on a constant energy surface, we have obtained a notion
of time along the extremum path of the particle.

Suppose we consider an arbitrary variation $\delta q$ about an
extremal path $\msf{q}(s)$ with no conditions on the variation at the
beginning and end points of the path. The variation $\delta W$ is then
purely the boundary term in \eqref{Wvartau}:
\begin{equation}
  \delta W  =  g_{ik} \frac{\ud q^i}{\ud \tau}\delta q^k\bigg|_{\tau_0}^{\tau_1}\ .  
\end{equation}
Clearly, the above equation implies that the partial derivative of $W$
with respect to the endpoint $q^i = \msf{q}^i(\tau_1)$ is
\begin{equation}
  \frac{\partial W}{\partial q^i} = g_{ij} \frac{\ud q^j}{\ud \tau}\ ,
\end{equation}
which agrees with the expression \eqref{Wtangder} once we use
$\ud W / \ud s_1 = \sqrt{2(E-V)}$ and \eqref{staudef}. Recall from
Hamilton-Jacobi theory that the expression for the conjugate momentum
$p_i$ is the partial derivative of $W$ with respect to $q^i$. Thus,
\begin{equation}
  p_i = g_{ij}\frac{\ud q^j}{\ud \tau}\ .
\end{equation}
It is satisfying to see that we recover the usual expression for
momentum in terms of the particle velocity with respect to the new
time $\tau$ \eqref{staudef}.

The definition of $\tau$ \eqref{staudef} in terms of dynamics in
configuration space is the central equation of our paper and arises in
any system that satisfies a Hamilton-Jacobi equation
\eqref{timeindHJ}. This origin of the parameter $\tau$ allows us to
extend its interpretation for classically forbidden paths in
configuration space, viz., paths for which $E < V$, as emphasized in
\cite{Wadia:1979yu}. In this case, the ansatz for the wavefunction
$\psi(q)$ is $\psi(q) = \e^{-W(q) / \hbar}\chi(q)$. The Schrodinger
equation for a particle in one dimension at leading order in $\hbar$
then becomes
\begin{equation}\label{hamjacgeninv}
	-\frac{1}{2m}  \left( \frac{\ud W}{\ud q}\right)^2 + V(q) = E\ .
\end{equation}
which is consistent with $E < V$. The analogous equation for the
particle on the Riemannian manifold is
\begin{equation}\label{hamjacgeninv1}
	-\frac{1}{2}  g^{ij}(q)  \frac{\partial W}{\partial q^i} \frac{\partial W}{\partial q^j} + V(q) = E\ ,
\end{equation}
which can be solved in the same way as earlier, giving an equation for
the semiclassical tunnelling path:
\begin{equation}\label{eomtunnel}
	\frac{\ud^2 \tl\mq^i}{\ud \tau^2} + \Gamma^{i}_{jk} \frac{\ud \tl\mq^j}{\ud \tau}\frac{\ud \tl\mq^k}{\ud \tau} - g^{ij}\frac{\partial V}{\partial \tl\mq^j} = 0\ .
\end{equation}
Note that the sign of the potential term is flipped compared to the
classically allowed case \eqref{eom}. This has the interpretation that
the semiclassical tunnelling path can be understood as a classically
allowed path for a particle in the inverted potential $\tl{V} = -V$
with energy $\tl{E} = -E$ (this is consistent since
$\tl{E} - \tl{V} = -(E-V) > 0$ which is indeed the allowed value of $\tl{E}$
for a classically allowed path for a particle in the inverted
potential $\tl{V}$).

\section{Non-Abelian Gauge Theory}

In this section, we consider a situation in which the dynamics in
configuration space occurs in the presence of a gauge symmetry. The
prototypical example of this is gauge field theory where the gauge
symmetry is based on a compact, semi-simple group $G$. We show that
there is a natural way to incorporate the gauge symmetry into our
previous analysis, and we again arrive at a definition of time based
on dynamics in the configuration space of gauge fields, but now modulo
the gauge symmetries. The techniques of this section carry over to the
analysis for general relativity as well, where the gauge symmetry is
that of spatial diffeomorphisms.

\subsection{The Hamilton-Jacobi equation for gauge theory}

The Hamilton-Jacobi method in $3+1$ dimensional Yang-Mills theory was
originally discussed in \cite{Wadia:1979yu} to understand the meaning
of instantons -- classical solutions of euclidean Yang-Mills equations
-- in the Schrodinger picture of quantum mechanics. In the
semi-classical limit, these serve as dominant tunnelling
configurations between Yang-Mills vacua of different winding
number. In the Hamiltonian approach, the degrees of freedom of
Yang-Mills theory with compact semi-simple gauge group $G$ are the
gauge fields $A_i^a$, on a three dimensional surface $\Sigma$ with
euclidean metric $\delta_{ij}$. The index $a$ runs over the $\dim\, G$
basis of the Lie algebra, with the Killing form
$\mc{K}^{ab} = \delta^{ab}$ used to raise and lower Lie algebra
indices. The conjugate momentum corresponding to $A_i^a(x)$ is
$\pi^{i}_a(x)$ with the Poisson bracket
$\{A_i^a(x), \pi^j_b(y) \} = \delta_i^j \delta^a_b \delta(x-y)$. The
Hamiltonian is
\begin{align}\label{hamYM}
	H &= \frac{1}{2}\int_\Sigma \ud^{3}x\,(\delta^{ab} \delta_{ij}\pi^i_a \pi^j_b) + V[A_i^a]\ ,\quad\text{with}\quad V[A_i^a] = \frac{1}{4}\int_\Sigma\ud^{3}x\, F_{ij}^a F^{ij,a}\ .
\end{align}
There is a constraint on the phase space which is the Gauss' law:
\begin{equation}\label{gausslaw}
	\nabla_i \pi^{i,a} = 0\ ,
\end{equation}
where $\nabla_i = \partial_i + \i g A_i^a T^a$ is the covariant
derivative associated to the gauge field $A_i^a$. Since we are
interested in semiclassical tunneling solutions between vacua of the
Yang-Mills theory (so, the energy $E = 0$), we take the wavefunction
to be of the form $\Psi[A_i^a(x)] = \e^{-W[A_i^a(x)]/ \hbar}$ and plug
it into the Schrodinger equation $H \Psi[A_i^a(x)] = 0$:
\begin{equation}
	\left(-\frac{\hbar^2}{2}\int_\Sigma \ud^3x\, \frac{\delta}{\delta A_i^a(x)}\frac{\delta}{\delta A_i^a(x)} + V[A_i^a]\right) \e^{-W[A_i^a]/ \hbar} = 0\ .
\end{equation}
The leading term in the $\hbar \to 0$ limit is
\begin{equation}\label{HJgauge}
	-\frac{1}{2}\int_\Sigma \ud^{3}x\, \frac{\delta W}{\delta A_i^a(x)} \frac{\delta W}{\delta A_i^a(x)} +   V[A_i^a] =  0\ ,
\end{equation}
which is the Hamilton-Jacobi equation for the non-abelian gauge
theory. The Gauss' law constraint \eqref{gausslaw} gives
\begin{equation}\label{gaugecons}
  \nabla_i \frac{\delta W}{\delta A_i^a} = 0\ .
\end{equation}
The meaning of the above constraint is clear if we multiply the above
equation by a gauge transformation parameter $\eta^a(x)$ and
integrating over $\Sigma$:
\begin{equation}\label{gausslawdemo}
	0 =   \int_\Sigma\ud^{3}x\, \eta^a(x) \nabla_i \frac{\delta W}{\delta A_i^a} = - \int_\Sigma\ud^{3}x\, \nabla_i\eta^a(x) \frac{\delta W}{\delta A_i^a} = -\int_\Sigma\ud^{3}x\, \delta_{\eta} W\ ,
\end{equation}
where $\delta_\eta W$ is the gauge transformation of $W$ with
parameter $\eta^a$. We have discarded the boundary term that arises
from integrating by parts in the second step by considering only those
$\eta^a$ which are zero on the boundary, the so-called \emph{small
  gauge transformations}. Thus, the Gauss' law constraint tells us
that $W$ must be invariant under small gauge transformations. Keeping
the above in mind, let us look at infinitesimal deformations of the
form $\delta A_i^a - \nabla_i\delta \eta_a$ in the configuration space
of gauge fields. There is a natural positive-definite, gauge invariant
metric on these deformations that can be inferred from the kinetic
term in the Hamiltonian \eqref{hamYM}:
\begin{equation}\label{gaugemet}
  \ud s^2 = \int_\Sigma \ud^{3}x\, \delta_{ab}\,\delta^{ij}\, \big(\delta A^a_i(x) - \nabla_i \delta \eta^a(x)\big)\big(\delta A^b_j(x) - \nabla_j \delta \eta^b(x)\big)\ .
\end{equation}
As in the particle mechanics example, the Hamilton-Jacobi equation
\eqref{HJgauge} is solved by choosing $W[A^a_i(x)]$ to be the value of
a functional $W[A^a_i(x,s)]$ on an extremum path $A_i^a(x,s)$,
$s_0 \leq s \leq s_1$, in configuration space, where $s$ is the
distance along the path as measured by the metric
\eqref{gaugemet}.\footnote{We use the same notation for both the path
  variable $A_i^a(x,s)$ and the end-point value
  $A_i^a(x) = A_i^a(x,s_1)$ to avoid excessive notation.} Since the
path is extremal, the functional derivative of $W[A_i^a]$ that appears
in the Hamilton-Jacobi equation \eqref{HJgauge} will be non-zero only
along the tangent vector to the classical path:
\begin{equation}\label{Wtangdergauge}
  \frac{\delta W}{\delta A_i^a(x)} = \delta_{ab} \delta^{ij} \left(\frac{\ud A_{j}^{b}}{\ud s} - \nabla_j \frac{\ud \eta^b}{\ud s}\right)\frac{\ud W}{\ud s}\bigg|_{s=s_1}\ ,
\end{equation}
where the factor $\delta_{ab}\delta^{ij}$ is the metric on tangent
deformations given in \eqref{gaugemet}. Just as in the particle
mechanics case, the tangent vector
$\frac{\ud A^a_i}{\ud s} - \nabla_i \frac{\ud \eta^a}{\ud s}$ has unit
norm with respect to the metric \eqref{gaugemet}. Plugging the
steepest descent expression \eqref{Wtangdergauge} into the
Hamilton-Jacobi equation \eqref{HJgauge}, we get
\begin{equation}\label{HJgaugesimple}
  \frac{1}{2}\left(\frac{\ud W}{\ud s_1}\right)^2 = V[A_i^a]\ ,
\end{equation}
with $V[A_i^a]$ given by \eqref{hamYM}.
Plugging in \eqref{Wtangdergauge} into the Gauss' law constraint gives
\begin{equation}\label{tangveccons}
  \nabla_i (A'^{b}_{i} - \nabla_i \eta'^b) = 0\ ,
\end{equation}
where we have denoted derivatives with respect to $s$ by a $'$ to
avoid clutter:
\begin{equation}
  A'^a_i \equiv \frac{\ud A^a_i}{\ud s}\ ,\quad   \eta'^a \equiv \frac{\ud \eta^a}{\ud s}\ .
\end{equation}
The expression for $W$ that follows from \eqref{HJgaugesimple} is then
\begin{equation}\label{Sintexpr}
	W[A_i^a, \eta^a, \alpha^a] = \int_{s_0}^{s_1} \ud s\,\sqrt{2V[A_i^a]} + \int_{s_0}^{s_1} \ud s\, G[\alpha^a,\eta^a,A_i^a]\ ,
\end{equation}
with
\begin{equation}
	G[\alpha^a, \eta^a, A_i^a] = \int_\Sigma \ud^{3}x\,\alpha^a \nabla_i( A'^a_i - \nabla_i \eta'^a)\ ,
\end{equation}
where $\alpha^a$ is the Lagrange multiplier field which imposes the
Gauss' law constraint in \eqref{tangveccons}. The line element $\ud s$
appearing in \eqref{Sintexpr} is \eqref{gaugemet}
\begin{align}\label{orbitmet}
  \ud s^2 &=  \int_\Sigma \ud^{3}x\, \left(\delta A_i^a - \nabla_i \delta \eta^a\right) \left(\delta A_i^a - \nabla_i \delta\eta^a\right)\nonumber\\
          &=  \int_\Sigma \ud^{3}x\, \left(\delta A_i^a \delta A_i^a -  \nabla_i\delta \eta^a \nabla_i \delta \eta^a\right)\ ,
\end{align}
where $\delta A^a_i$ and $\delta \eta^a$ are tangent deformations
along the path, and the second formula is obtained by using
\eqref{tangveccons}.

\subsection{Deriving the Yang-Mills equations}
In this subsection, we extremize Hamilton's characteristic function
$W$ in \eqref{Sintexpr}. The path variation of $W$ is
\begin{equation}
  \hvar W = \int_{s_0}^{s_1} (\hvar\ud  s)\left(\sqrt{2V} + G\right) +  \int_{s_0}^{s_1} \ud s \left(\hvar \sqrt{2V} + \hvar G\right)\ ,
\end{equation}
The various variations are
\begin{align}\label{EVvar}
	\hvar \sqrt{2V} &= \frac{1}{\sqrt{2V}}\int_\Sigma \ud^{3}x\, (\nabla_j \hvar A^a_i) F^{ji,a}\ ,\nonumber\\
	&= \frac{1}{\sqrt{2V}} \left(-\int_\Sigma \ud^{3}x\, \hvar A^a_i \nabla_j F^{ji,a} + \int_{\partial\Sigma} \ud^{2}\sigma_j\, \hvar A^a_i F^{ji,a}\right)\ ,
\end{align}
\begin{align}\label{Gvar}
	\hvar G
	&= \int_\Sigma \ud^{3}x\Big(  \hvar \alpha^a \nabla_i ({A}'^a_i - \nabla_i \eta'^a) - \hvar \nabla_i \alpha^a (A'^a_i - \nabla_i\eta'^a) - \nabla_i\alpha^a \hvar(A'^a_i - \nabla_i\eta'^a)\Big) \nonumber\\
	&\quad + \int_{\partial\Sigma} \ud^{2}\sigma_i\, \alpha^a \hvar(A'^a_i - \nabla_i\eta'^a)\ ,
\end{align}
where $\hvar \nabla_i$ is the change in the covariant derivative that
results from a change in the gauge field,
$\hvar \nabla_i \alpha^a = \i g [\hvar A_i, \alpha]^a = \i g f^{abc} \hvar
A_i^b \alpha^c$. The path-variation $\hvar \ud s$ is
\begin{align}\label{dsvargauge}
	\hvar \ud s &= \ud s\,  \int_\Sigma \ud^{3} x\, (\hvar {A}'^a_i {A}'^a_i - \hvar \nabla_i {\eta}'^a \nabla_i{\eta}'^a - \nabla_i \hvar {\eta}'^a \nabla_i{\eta}'^a)\ .
\end{align}
We again introduce the parameter $\tau$ defined via
\begin{equation}\label{staudef1}
	\frac{\ud}{\ud\tau} =  \sqrt{2V} \frac{\ud}{\ud s}\ ,\quad  \ud \tau = \frac{\ud s}{\sqrt{2V}}\ ,
\end{equation}
which is analogous to the equation \eqref{staudef} in the particle
mechanics case. We express all quantities in terms of derivatives
with respect to $\tau$, which we denote by a dot above the
quantity:
\begin{equation}
  \dot A^a_i \equiv \frac{\ud A^a_i}{\ud \tau}\ ,\quad   \dot\eta^a \equiv \frac{\ud \eta^a}{\ud \tau}\ .
\end{equation}
First, extremizing with respect to $\hvar \eta^a$ gives
\begin{align}\label{hvarlambda}
	0 &= \int \ud \tau  \int_\Sigma \ud^{3}x\,(\nabla_i\alpha^a \nabla_i \hvar \dot \eta^a - \nabla_i \hvar \dot\eta^a \nabla_i\dot\eta^a ) = \int \ud \tau  \int_\Sigma \ud^{3}x\, \nabla_i(\alpha^a - \dot \eta^a) \nabla_i \hvar \dot \eta^a\ .
\end{align}
This gives the equation
\begin{equation}
  \nabla_i\nabla_i (\alpha^a - \dot\eta^a) = 0\ .
\end{equation}
Since $\nabla_i\nabla_i$ is a positive definite operator on the gauge
parameters which decay to zero sufficiently rapidly at infinity, the
only solution to the above is $\alpha^a = \dot \eta^a$. We replace
every occurrence of $\dot\eta^a$ with $\alpha^a$ here onwards.

Extremizing with respect to $\alpha^a$ gives the constraint
\begin{equation}\label{gausseq}
  \nabla_i(\dot A^a_i - \nabla_i \alpha^a) = 0\ .
\end{equation}
Identifying $\tau$ as the time direction of euclidean spacetime, it is
clear that the combination $\partial_\tau A_i^a - \nabla_i \alpha^a$
plays the role of the electric field. Indeed, identifying $\alpha^a$
with the time component of the spacetime gauge field $A^a_\tau$, the
combination $\partial_\tau A_i^a - \nabla_i \alpha^a = F_{\tau i}^a$
is precisely the electric field (which is the canonically conjugate
momentum $\pi^a_i$). The above equation is the $\tau$ component of the
euclidean Yang-Mills equations $\nabla_\mu F^{\mu\nu} = 0$ with
$\nu = \tau$:
\begin{equation}\label{YMtimecomp}
  \nabla_i F^{i\tau,a} = 0\ .
\end{equation}
The variation with respect to $A_i^a$ has the following terms:
\begin{align}
	\int\ud \tau\int_\Sigma \ud^{3}x\, \Big( & \hvar  \dot A^a_i  (\dot A^a_i - \nabla_i \alpha^a)- \i g f^{abc} \hvar A^a_i \alpha^b (\dot A_i^c - \nabla_i \alpha^c) - \hvar A_i^a \nabla_j F^{ji,a} \Big)\ .
\end{align}
The bulk equation of motion is then
\begin{equation}
	-\partial_\tau\left(\partial_\tau A_i^a - \nabla_i \alpha^a\right) - \i g f^{abc} \alpha^b\left(\partial_\tau A_i^c - \nabla_i \alpha^c\right) - \nabla_j F^{ji,a} = 0\ .
\end{equation}
which are the spatial components of the Yang-Mills equations
$\nabla_\mu F^{\mu\nu}=0$ with $\nu = i$:
\begin{equation}\label{YMeq}
	\nabla_\tau F^a_{\tau i} + \nabla_j F^{ji,a} = 0\quad\text{that is,}\quad \nabla_\tau F^{\tau i,a} + \nabla_j F^{ji,a} = 0\ .
\end{equation}

\subsection{Boundary terms}
We now discuss the boundary integrals that occur at various stages of
our computation. They are present in \eqref{EVvar} and \eqref{Gvar},
and we collect them below:
\begin{align}\label{gaugebdryterms}
  \int \ud\tau \int_{\partial\Sigma} \ud^{2}\sigma\, r_i \left( \hvar A_j^a F^{ji,a} + A_\tau^a \hvar \pi^a_i\right)\ .
\end{align}
To ensure a good variational principle, one must make sure that there
are no boundary terms proportional to the variations of the
fields. The above boundary terms go to zero when we impose the
boundary conditions
\begin{equation}\label{gaugebc}
	\text{On $\partial\Sigma$}:\quad  \hvar A_i^a = 0\ ,\quad r^i \hvar \pi_i^a = 0\ ,
\end{equation}
where $r^i$ is the unit normal to $\partial\Sigma$. The gauge field
$A^a_i$ and the normal component of the electric field $\pi^a_i$
satisfies Dirichlet boundary conditions. One can formulate the `dual'
variational problem by adding the following boundary term to
$G[A^a_i]$:
\begin{equation}\label{gaugebcham}
  -\int_{\partial\Sigma} \ud^{2}\sigma_i\, A_\tau^a \pi^{i,a}\ ,
\end{equation}
which modifies the boundary terms to
\begin{align}\label{gaugebdryham}
  \int \ud\tau \int_{\partial\Sigma} \ud^{2}\sigma\, r_i \left( \hvar A_j^a F^{ji,a} - \hvar A_\tau^a \pi^{i,a}\right)\ .
\end{align}
Now, the above boundary terms can be eliminated by setting
$\hvar A^a_\tau = 0$ and $\hvar A^a_i = 0$ on $\partial\Sigma$. This
choice of boundary conditions corresponds to the usual Dirichlet
boundary conditions that one imposes on all components of the gauge
field in the Lagrangian formulation of the theory. Indeed, adding the
term \eqref{gaugebcham} to $G$ finally results in the Hamiltonian $H$
that one obtains from the Lorentz covariant Lagrangian of the theory.

\section{General Relativity}\label{GRsec}

General relativity describes gravitational physics in $d+1$ spacetime
dimensions in terms of a $d+1$ dimensional manifold with a Lorentzian
signature metric on it. We assume that the $d+1$ dimensional manifold
can be foliated by $d$ dimensional hypersurfaces $\Sigma$ which are
spatial with respect to the $d+1$ dimensional metric. We restrict to
$d \geq 2$ here. The Hamilton-Jacobi approach starts with recognizing
that the dynamical variables are the components of the metric $g_{ij}$
on the $d$ dimensional manifold $\Sigma$, so that the configuration
space is $\mc{M}_\Sigma$ -- the space of metrics on $\Sigma$.

\subsection{The Einstein-Hamilton-Jacobi equations}

The Einstein-Hamilton-Jacobi equations are partial differential
equations on $\mc{M}_\Sigma$ for the Hamilton's principal function
$S[g_{ij}(x)]$ which is a functional on $\mc{M}_\Sigma$. These were
first written by Peres \cite{Peres1962OnCP}:
\begin{align}\label{ehj}
  &\cG_{ijkl}(x) \frac{\delta S}{\delta g_{ij}(x)} \frac{\delta S}{\delta g_{kl}(x)} - \sqrt{g} \big(R(x) - 2\Lambda\big) = 0\ ,\\
  &D_i \frac{\delta S}{\delta g_{ij}(x)} = 0\ .\label{3diff}
\end{align}
where (1) $g$ is the determinant of $g_{ij}$, (2) $D_i$ is the
covariant derivative compatible with $g_{ij}(x)$, (3) $R$ is the Ricci
scalar of $g_{ij}$, (4) $\Lambda$ is the cosmological constant, and
(5) $\cG_{ijkl}$ are components of the inverse de Witt metric
\begin{equation}\label{invdeWitt}
  \cG_{ijkl} = \frac{1}{2 \sqrt{g}} \left(g_{ik} g_{jl} + g_{il} g_{jk} - \frac{2}{d-1} g_{ij} g_{kl}\right)\ ,
\end{equation}
with the de Witt metric $\cG^{ijkl}$ \cite{DeWitt:1967yk} itself being
\begin{equation}\label{deWitt}
  \cG^{ijkl}(x) =  \frac{1}{2} \sqrt{g} (g^{ik} g^{jl} + g^{il} g^{jk} - 2 g^{ij} g^{kl})\ .
\end{equation}
The Einstein-Hamilton-Jacobi equation \eqref{ehj} resembles the
Hamilton-Jacobi equations in the particle mechanics and gauge theory
examples, except that it is local on $\Sigma$. Before we embark on
solving it, we consider the second set of equations
\eqref{3diff}. These implement the $d$-diffeomorphism invariance of
Hamilton's principal function $S$, which can be seen as
follows. Consider a vector field $\xi^i$ on $\Sigma$ which vanishes on
the boundary $\partial\Sigma$.\footnote{The analogous statement for
  asymptotic regions is that the vector field dies sufficiently
  rapidly as asymptotic infinity is approached.} Then,
\begin{equation}\label{Sdiffinv}
D_i \frac{\delta S}{\delta g_{ij}} = 0\ \Rightarrow\ \int_\Sigma \ud^{d}x\, 2\xi_i D_j\frac{\delta S}{\delta g_{ij}} = -\int_\Sigma\ud^{d}x\,  2D_i \xi_j \frac{\delta S}{\delta g_{ij}} = -\int_\Sigma\ud^{d}x\,  2D_{(i} \xi_{j)}\, \frac{\delta S}{\delta g_{ij}} = 0\ ,
\end{equation}
where the boundary integral that arises in the integration by parts in
the second step drops out since $\xi^i$ vanishes on the
boundaries. Such $\xi^i$ generate the so-called \emph{small}
diffeomorphisms and the above computation shows that $S$ is invariant
under them.

\subsection{An expression for Hamilton's principal function}

The equation \eqref{ehj} can be solved as in the earlier sections
by interpreting the Hamilton's principal function $S[g_{ij}(x)]$ as
the value of a functional $S[g_{ij}(x,\lambda)]$ on an extremum path
$g_{ij}(x,\lambda)$ in the configuration space of metrics
$\mc{M}_\Sigma$, with path parameter $\lambda$. Along a path, the equation \eqref{ehj} becomes
\begin{align}\label{ehjpath}
  &\cG_{ijkl}(x,\lambda) \frac{\delta S}{\delta g_{ij}(x,\lambda)} \frac{\delta S}{\delta g_{kl}(x,\lambda)} - \sqrt{g}(x,\lambda) \big(R(x,\lambda) - 2\Lambda\big) = 0\ .
\end{align}
There is one equation for each point on $\Sigma$, and for each point
on the path. However, to apply the methods of the previous sections,
it is useful to convert the above local equation into one single
equation at each point of the path. This can be done while still
retaining the locality of \eqref{ehjpath} as follows. Suppose, at each
point $\lambda$ of the path, we smear the equation over $\Sigma$ with
a strictly positive -- but otherwise arbitrary -- smearing function
$N(x,\lambda)$:
\begin{align}\label{ehjsmear}
  &\int_\Sigma \ud^{d}x\,  N\cG_{ijkl} \frac{\delta S}{\delta g_{ij}} \frac{\delta S}{\delta g_{kl}} -  \int_\Sigma \ud^{d}x\,  N\sqrt{g} \big(R - 2\Lambda\big) = 0\ .
\end{align}
Since $N(x,\lambda)$ is arbitrary, it is possible get back the local
equations by considering $N(x,\lambda)$ which are supported only in an
infinitesimal neighbourhood of any given point on $\Sigma$. We call
$N(x,\lambda)$ the \emph{lapse} function in anticipation of the role
it will eventually play.

Analogous to our discussion in the gauge theory example (see the
paragraphs around \eqref{gaugemet}), we take the infinitesimal
deformation in the configuration space of metrics to be
$\delta g_{ij} - 2 D_{(i} \delta M_{j)}$ where the second term is an
infinitesimal $d$-diffeomorphism with parameter $\delta M^j$. The
corresponding `smeared' de Witt metric is
\begin{align}\label{deWittsmear}
\ud s^2 =  \int_\Sigma \ud^{d}x\, N^{-1} \cG^{ijkl} \big(\delta g_{ij}(x) - 2 D_{(i} \delta M_{j)}(x)\big)\big(\delta g_{kl}(x) -  2 D_{(k} \delta M_{l)}(x)\big)\ .
\end{align}
Note that the above line element can be positive, negative or zero
since the de Witt metric has indefinite signature.  This
indefiniteness can be exhibited by decomposing the metric in terms of
the conformal mode $\Omega(x)$ and the rest as (for instance, see
\cite[Eq.(5.7)]{DeWitt:1967yk}):
\begin{equation}\label{metdecomp}
  {g}_{ij}(x) = g(x)^{1/d}\, \tl{g}_{ij}(x)\ ,\quad \Omega (x) = g(x)^{1/4}\ ,
\end{equation}
where $g$ is the determinant of $g_{ij}$. 
In terms of $\Omega$ and $\tl{g}_{ij}$ the de Witt metric
\eqref{deWitt} clearly exhibits indefiniteness due to the negative
signature of the deformations along the conformal mode:
\begin{equation}\label{deWittdecomp}
  \cG^{ijkl}(x) \delta g_{ij}(x) \delta g_{kl}(x) = -\frac{16(d-1)}{d} \delta \Omega(x)^2 + \Omega^2(x)\, \tl{g}^{ij} \delta \tl{g}_{jk}(x)\, \tl{g}^{kl} \delta \tl{g}_{li}(x)\ .
\end{equation}
As earlier, we can parametrize the path by the distance measured with
respect to the smeared de Witt metric \eqref{deWittsmear} while taking
into account its indefinite signature. Let $\epsilon = \sign(\ud s^2)$
along the path (with $\epsilon = 0$ when $\ud s^2 = 0$), and suppose
we consider a portion of the path where $\epsilon$ is fixed to one
value. Let us define the following quantity which is always positive
or zero along that portion:
\begin{equation}\label{sepsdef}
  \ud s_\epsilon^2 = \epsilon \times \ud s^2 = |\ud s^2|\ .
\end{equation}
We can then parametrize the path with $s_\epsilon$, the integral of
$\ud s_\epsilon = \sqrt{|\ud s^2|}$ along the path -- as long as
$\epsilon \neq 0$.\footnote{It would be interesting to have a
  classification of paths, if possible, in the configuration space of
  metrics $\mc{M}_\Sigma$ with a particular value of $\epsilon$. In
  this case, one can restrict the subsequent steps below to a single
  class with a particular value of $\epsilon$. For instance, paths for
  which the tangent vector $\ud g_{ij} / \ud s_\epsilon$ is traceless
  have $\epsilon = +1$, simply because the term $-2g^{ij} g^{kl}$ in
  the de Witt metric \eqref{deWitt} that is responsible for the
  indefinite signature drops out for such tangent vectors.}

Since the path is extremal, i.e., of steepest descent, in the
configuration space $\mc{M}_\Sigma$, the functional derivative
$\delta S / \delta g_{ij}$ is non-zero only along the tangent vector
to the path:
\begin{equation}\label{tangexpr}
  \frac{\delta S}{\delta g_{ij}} = N^{-1} \cG{}^{ijkl} \left(\frac{\ud g_{ij}}{\ud s_\epsilon} - 2 D_{(i} \frac{\ud M_{j)}}{\ud s_\epsilon}\right) \frac{\ud S}{\ud s_\epsilon}\ .
\end{equation}
As in the previous sections, it is convenient to denote derivatives
with respect to $s_\epsilon$ with a $'$:
\begin{equation}
  \frac{\ud g_{ij}}{\ud s_\epsilon} = g'_{ij}\ ,\quad   \frac{\ud M_{j}}{\ud s_\epsilon} = M'_j\ .
\end{equation}
From the definition of $\ud s_\epsilon$ \eqref{sepsdef}, and the metric
\eqref{deWittsmear}, it is clear that the tangent vector
$\frac{\ud g_{ij}}{\ud s_\epsilon} - 2 D_{(i} \frac{\ud M_{j)}}{\ud
  s_\epsilon}$ has norm $\epsilon$ with respect to
\eqref{deWittsmear}:
\begin{equation}\label{epsdef}
  \epsilon =   \int_\Sigma \ud^{d}x\, N^{-1} \cG^{ijkl} \big( g'_{ij} - 2 D_{(i} \delta M'_{j)}\big)\big( g'_{kl} -  2 D_{(k}  M'_{l)}\big)\ .
\end{equation}
Plugging in the expression \eqref{tangexpr} into the smeared
Einstein-Hamilton-Jacobi equation \eqref{ehjsmear}, we get
\begin{align}\label{ehjtangunit}
  & \epsilon\left(\frac{\ud S}{\ud s_\epsilon}\right)^2 = \int_\Sigma \ud^{d}x\, N \sqrt{g} \big(R - 2\Lambda\big)\ .
\end{align}
The $d$-diffeomorphism constraint \eqref{3diff} becomes
\begin{equation}\label{momcons}
  D_i \left(N^{-1} \cG^{ijkl} (g'_{kl} - 2D_{(k} M'_{l)})\right) = 0\ .
\end{equation}
\textbf{Note:} When the norm $\epsilon$ \eqref{epsdef} is zero, the
left hand side of \eqref{ehjtangunit} vanishes and the equation
\eqref{ehjtangunit} does not determine $S$; we cannot proceed with
the Hamilton-Jacobi analysis (see the end of this subsection for an
example). We thus restrict ourselves to the case of non-zero norm,
i.e., $\epsilon \neq 0$. As long as $\epsilon \neq 0$, the equation
\eqref{ehjtangunit} is non-trivial along the path, and one can proceed
with the computation as in the previous sections and derive an
equation for the classical path.

Thus, we get an expression for $S$ as a functional of the classical
path:
\begin{equation}\label{Sdef}
  S[g_{ij},M_i,N,N_i] = \int_{s_{\epsilon,0}}^{s_{\epsilon,1}} \ud s_\epsilon \left(\sqrt{-\epsilon V[g_{ij},N]} + \epsilon C[g_{ij},M_i,N,N_i]\right)\ ,
\end{equation}
where $g_{ij}(x,s_\epsilon)$, $N(x,s_\epsilon)$, $M_i(x,s_\epsilon)$
and $N_i(x,s_\epsilon)$ are defined on the classical path with
parameter $s_\epsilon$, and the functionals $V[g_{ij},N]$ and
$C[g_{ij},M_i,N,N_i]$ are given by
\begin{align}\label{VCdef}
  V[g_{ij},N] &= -\int_\Sigma\ud^{d}x\, \sqrt{g}  N (R-2\Lambda)\ ,\nonumber\\
  C[g_{ij},M_i,N,N_i] &= \int_\Sigma \ud^{d}x\,   N_{j} D_i\! \left(N^{-1} \cG^{ijkl}\big( g'_{kl} - 2 D_{k}  M'_{l}\big)\right)\ ,
\end{align}
where $N_i$ is a Lagrange multiplier field which implements
\eqref{momcons}. The infinitesimal distance $\ud s$ along the path 
is given by
\begin{align}\label{orbmet}
  \ud s_\epsilon^2 &=  \epsilon   \int_\Sigma \ud^{d}x\, N^{-1} \cG^{ijkl} \big(\delta g_{ij} - 2 D_{(i} \delta M_{j)}\big)\big(\delta g_{kl} -  2 D_{(k} \delta M_{l)}\big)\ ,\nonumber\\
                   &=   \epsilon    \int_\Sigma \ud^{d}x\, N^{-1} \cG^{ijkl} \big(\delta g_{ij}\delta g_{kl} -  4   D_{(i} \delta M_{j)} D_{(k} \delta M_{l)}\big)\ ,
\end{align}
where $\delta g_{ij} - 2 D_{(i} \delta M_{j)}$ is along the tangent
vector to the path, and the second line is obtained by using the
orthogonality \eqref{momcons}.

\textbf{Note:} The explicit factor of $\epsilon$ in the second term in
\eqref{Sdef} is not standard. We have inserted it so that the
equations of motion that we derive eventually \eqref{conseq},
\eqref{geq}, \eqref{pieq} do not depend on $\epsilon$. This step is
justified since we can absorb $\epsilon$ into the Lagrange multiplier
field $N_i$ by a redefinition as long as $\epsilon \neq 0$.

\paragraph{Comments on the sign $\epsilon$} 
The equation \eqref{ehjtangunit} implies that $\epsilon$ -- originally
defined to be the sign of the norm squared of the tangent vector
\eqref{epsdef} -- is also the sign of the quantity on the right hand
side of \eqref{ehjtangunit}:
\begin{equation}\label{epscons}
  \epsilon = \text{sign}\left(\int_\Sigma \ud^{d}x\, N \sqrt{g} (R- 2\Lambda)\right)\ .
\end{equation}
The sign $\epsilon$ of the norm of the tangent vector is a new
ingredient in the Hamilton-Jacobi analysis that is specific to general
relativity due to the indefiniteness of the de Witt metric. Recall the
decomposition of the de Witt metric \eqref{deWittdecomp} which we
reproduce here for convenience:
\begin{equation}\label{deWittdecomp1}
  \cG^{ijkl}(x) \delta g_{ij}(x) \delta g_{kl}(x) = -\frac{16(d-1)}{d} \delta \Omega(x)^2 + \Omega^2(x)\, \tl{g}^{ij} \delta \tl{g}_{jk}(x)\, \tl{g}^{kl} \delta \tl{g}_{li}(x)\ .
\end{equation}
It is clear that $\epsilon$ is $-1$ when the conformal mode
contribution in the first term dominates over the second term. We
could freeze the $\tl{g}_{ij}$ degrees of freedom in which case the
tangent vector is completely along the conformal mode:
$\ud \tl{g}_{ij} / \ud s_\epsilon = 0$. Then, the spatial metric
depends on $s$ only through the conformal mode
$g_{ij} = \Omega^{4/d}(s) \tl{g}_{ij}$. This possibility is realized
in de Sitter spacetime, as we discuss in Section \ref{examples}.

The opposite possibility, $\epsilon = +1$, arises when the
contribution of $\tl{g}_{ij}$ in the second term in
\eqref{deWittdecomp} outweighs that of the conformal mode. Again, one
can consider the extreme situation where the conformal mode is
altogether frozen $\ud \Omega / \ud s_\epsilon = 0$, i.e.,
$g^{ij} \frac{\ud g_{ij}}{\ud s_\epsilon} = 0$. As we shall see later,
this translates to the condition that the trace of the extrinsic
curvature $K_{ij}$ of $\Sigma$ is zero which is known as the
\emph{maximal slicing} condition. This condition is always possible to
achieve in asymptotically AdS spacetimes
\cite{Wittentalk,Witten:2022xxp,Chrusciel:2022cjz}.

The case $\epsilon = 0$, i.e., when the tangent vector has zero norm,
also appears in many interesting situations. Any static spacetime
i.e., a spacetime with $g'_{ij} = 0$, $M'_i = 0$, has zero tangent, so
that $\epsilon$ is trivially zero. For an example in which the metric
depends on the path parameter $\lambda$, consider the path with
$M_i = 0$ and $g_{ij}(x,\lambda) = \lambda^{2c_i} \delta_{ij}$, where
$c_i$, $i=1,\ldots,d$, are constants that satisfy
$c_1 + \cdots + c_d = c_1^2 + \cdots + c_d^2 = 1$. The tangent vector
is then $\frac{\ud g_{ij}}{\ud \lambda} = 2c_i \lambda^{- 1} g_{ij}$,
and has zero norm:
$\cG^{ijkl} \frac{\ud g_{ij}}{\ud \lambda} \frac{\ud g_{kl}}{\ud
  \lambda} = 0$. This path in the configuration space of metrics is
nothing but the Kasner solution of the vacuum Einstein's equations in
$d+1$ dimensions. As noted earlier, our subsequent analysis of the
Hamilton-Jacobi equation cannot be applied to situations with
$\epsilon = 0$.

\subsection{A notion of time along the extremum path}\label{newtau}

Recall the expression \eqref{Sdef} for $S$, with the assumption
$\epsilon \neq 0$:
\begin{equation}\label{Sdef1}
  S[g_{ij},M_i,N,N_i] = \int \ud s_\epsilon \left(\sqrt{-\epsilon V[g_{ij},N]} + \epsilon C[g_{ij},M_i,N,N_i]\right)\ ,
\end{equation}
with the functionals $V[g_{ij},N]$ and $C[g_{ij},M_i,N,N_i]$ given by
\begin{align}\label{VCdef1}
  V[g_{ij},N] &=  -\int_\Sigma\ud^{d}x\, \sqrt{g}  N (R-2\Lambda)\ ,\nonumber\\
  C[g_{ij},M_i,N,N_i] &= \int_\Sigma \ud^{d}x\,   N_{j} D_i\! \left(N^{-1} \cG^{ijkl}\big( g'_{kl} - 2 D_{k}  M'_{l}\big)\right)\ .
\end{align}
The equation of the path that extremizes $S$ can then be obtained by
setting to zero the variation of $S$ with respect to the variation of
the fields $g_{ij}$, $ N_i$, $ M_i$ and $N$ along the path. The total
variation of the action is
\begin{align}\label{varS}
  \hvar S &=  \int_{s_{\epsilon,0}}^{s_{\epsilon,1}} \ud s_\epsilon \left(\hvar\sqrt{-\epsilon V} + \epsilon\hvar C\right) + \int_{s_{\epsilon,0}}^{s_{\epsilon,1}} (\hvar \ud s_\epsilon)\left( \sqrt{-\epsilon V} + \epsilon C\right) \ ,\nonumber\\
          &=\int_{s_{\epsilon,0}}^{s_{\epsilon,1}} \frac{\ud s_\epsilon}{2\sqrt{-\epsilon V}} \epsilon\left( \hvar (- V) + 2\sqrt{-\epsilon V}\hvar C\right) + \int_{s_{\epsilon,0}}^{s_{\epsilon,1}} \frac{\ud s_\epsilon}{2\sqrt{-\epsilon V}}\,  \frac{1}{2}(2 \sqrt{-\epsilon V})^2 \frac{\hvar \ud s_\epsilon}{\ud s_\epsilon}\ .
\end{align}
In the second line, we have discarded the term with $C$ since it is
zero after extremizing with respect to $N_i$. We have
\begin{align}\label{Vvar}
  & \hvar (-V)\nonumber\\
  &= \int_\Sigma \ud^{d}x\,\sqrt{g} \Big(\hvar N (R-2\Lambda) + \hvar g_{ab}\big(D^a D^b N - g^{ab} D^c D_c N -N (R^{ab} - \tfrac{1}{2} g^{ab} R + \Lambda g^{ab})    \big) \Big)\nonumber\\
  &\quad + \int_{\partial\Sigma} \ud^{d-1}\sigma_a\, \cG^{abcd} (N   D_b  \hvar g_{cd} -  D_b N \hvar g_{cd})\ .
\end{align}
Next, we have
\begin{align}\label{Cvar}
  \hvar C =&  \int_\Sigma \ud^{d}x\, \Big(\hvar N_{j}\, D_i\big(N^{-1} \cG^{ijkl}(  g'_{kl} - 2 D_{k}   M'_{l})\big)\nonumber\\
           &\qquad\qquad  - \hvar D_i N_{j}\,N^{-1} \cG^{ijkl}(  g'_{kl} - 2 D_{k}  M'_{l}) - D_i N_{j} \hvar\big(N^{-1} \cG^{ijkl}( g'_{kl} - 2 D_{k}  M'_{l})\big)\Big)\nonumber\\
           & + \int_{\partial\Sigma} \ud^{d-1}\sigma_i\, N_{j}\, \hvar\big(N^{-1} \cG^{ijkl}(  g'_{kl} - 2 D_{k}  M'_{l})\big)\ ,
\end{align}
where $\hvar D_i$ is the change in the covariant derivative due to a
change in the metric $\hvar g_{ab}$. The variation of $\ud s_\epsilon$
that follows from \eqref{orbmet} is
\begin{align}\label{dsvar}
  \frac{\hvar \ud s_\epsilon}{\ud s_\epsilon} &=  \frac{1}{2}\epsilon \int_\Sigma \ud^{d}x\, \big(-\hvar N^{-1} \cG^{ijkl} + N^{-1} \hvar \cG^{ijkl}\big) \Big( g'_{ij}  g'_{kl}  -  4 D_{i}   M'_{j}  D_{k}  M'_{l} \Big)\nonumber\\
                            &\quad +  \epsilon\int_\Sigma \ud^{d}x\, N^{-1} \cG^{ijkl} \Big(  \hvar  g'_{ij}  g'_{kl}  - 4  D_{i}  \hvar   M'_{j} D_{k}   M'_{l} - 4 \hvar D_{i}  M'_{j} D_{k}  M'_{l} \Big)\ .
\end{align}
Note that every term in $(\hvar \ud s_\epsilon) / \ud s_\epsilon$
contains two $\ud / \ud s_\epsilon$ derivatives, every term in
$\hvar C$ contains a single $\ud / \ud s_\epsilon$, and $\hvar (-V)$
contains no $\ud / \ud s_\epsilon$ derivatives. Accordingly, in
\eqref{varS}, $\hvar C$ is accompanied by one power of
$2\sqrt{-\epsilon V}$ and $(\hvar \ud s)/\ud s$ by two powers of
$2 \sqrt{-\epsilon V}$. Further, the measure along the path in
\eqref{varS} always appears in the combination
$\ud s_\epsilon / 2\sqrt{-\epsilon V}$. Thus we can define a new
parameter $\tau$ along the path by\footnote{There is an additional
  $\sqrt{2}$ compared to the definition \eqref{staudef} in classical
  mechanics since there is an extra factor of $1/2$ in the $p^2$ term
  in the Hamilton-Jacobi equation in classical mechanics compared to
  the Einstein-Hamilton-Jacobi equation \eqref{ehj}.}
\begin{equation}\label{dtau}
  2\sqrt{-\epsilon V} \frac{\ud}{\ud s_\epsilon} = \frac{\ud}{\ud \tau}\ ,\quad   \ud \tau = \frac{\ud s_\epsilon}{2\sqrt{-\epsilon V}}\ .
\end{equation}
The parameter $\tau$ given by
\begin{equation}
  \tau(s_\epsilon) = \int^{s_\epsilon} \frac{\ud \tl{s}_\epsilon}{2\sqrt{-\epsilon V}}\ ,
\end{equation}
defines a new notion of time along the extremum path in configuration
space. Note that the above definition of time is invariant under
$d$-diffeomorphisms.

\subsection{Einstein's equations}\label{einsteineq}

We next carry out the extremization of $S$ by isolating the
coefficients of the variations $\hvar N_i$, $\hvar M_i$, $\hvar N$ and
$\hvar g_{ij}$. We recast the expressions in $\hvar S$ terms of the
new time parameter $\tau$ using \eqref{dtau}. We denote a derivative
with respect to $\tau$ by a dot:
\begin{equation}
  \dot f \equiv  \frac{\ud f}{\ud \tau}\ .
\end{equation}

\paragraph{The $\hvar M_i$ equation} The terms involving
$\hvar M_i$ in $\hvar S$ \eqref{varS} come from \eqref{dsvar} and
\eqref{Cvar}, and can be written as
\begin{align}\label{Meom}
  0 
  &= \epsilon\int \ud \tau \int_\Sigma \ud^{d}x\, N^{-1} \cG^{ijkl}2 D_i \big(N_j -  \dot M_j\big) \,D_{k}\hvar  \dot M_{l}\ .
\end{align}
This gives the equation
\begin{equation}\label{Mieom}
  D_k \big(N^{-1} \cG^{ijkl} 2 D_i (N_j - \dot M_j)\big) = 0\ ,
\end{equation}
so that
\begin{equation}\label{NMXrel}
  N_j = \dot M_j + X_j\ ,
\end{equation}
where $X_j$ is a solution of the equation
$ D_k \big(N^{-1} \cG^{ijkl} 2 D_i X_j\big) = 0$.\footnote{Contracting
  the free index $l$ with an arbitrary small diffeomorphism parameter
  $\xi_l$ and integrating over $\Sigma$, we get
\begin{equation}
  \int_\Sigma\ud^dx\, N^{-1} \cG^{ijkl} 2 D_i X_j D_k \xi_l = 0\ ,\quad\text{for all $\xi_i$}\ ,\nonumber
\end{equation}
which is the statement that the diffeomorphism $D_{(i} X_{j)}$ is
`orthogonal' to all small diffeomorphisms. Such diffeomorphisms are
null with respect to $\cG^{ijkl}$, which are allowed in principle
since $\cG^{ijkl}$ is an indefinite metric. Note that if we are able
to restrict ourselves to the situation where the tangent deformations
$\dot g_{ij} - 2 D_{(i} \dot M_{j)}$ are all traceless with respect to
$g_{ij}$ (which corresponds to zero mean curvature $K = 0$, see
\eqref{Ktrdef} below), the de Witt metric restricted to this subspace
is positive definite. In this case, the only solution is $X_i = 0$
(Killing vectors are also solutions, but these are ruled out since
$X_i$ generates a small diffeomorphism). See Section \ref{adsexample}
for a situation where it always possible to set $K = 0$.}

\paragraph{The $\hvar N_i$ equation} This is simply the
$d$-diffeomorphism constraint \eqref{momcons}:
\begin{equation}\label{momcons1}
  D_i \left(N^{-1} \cG^{ijkl} (\dot g_{kl} - 2D_{(k} \dot M_{l)})\right) = 0\ .
\end{equation}

\paragraph{The $\hvar N$ equation} The equation of motion that follows
from extremizing with respect to $\hvar N$
\begin{equation}
 -\frac{1}{4 N^2} \cG^{ijkl} (\dot g_{ij} \dot g_{kl} - 4 D_i \dot M_j D_k \dot M_l) + \sqrt{g} (R - 2 \Lambda)  + \frac{1}{N^2} \cG^{ijkl} D_i N_j (\dot g_{kl} - 2 D_k \dot M_l) = 0\ ,
\end{equation}
which, upon substituting $N_j = \dot M_j + X_j$ from \eqref{NMXrel},
becomes
\begin{multline}\label{Neom}
  -\frac{1}{4 N^2}  \cG^{ijkl} (\dot g_{ij} - 2 D_i \dot M_j) (\dot g_{kl} - 2 D_k \dot M_l) + \sqrt{g} (R - 2 \Lambda)   + \frac{1}{N^2} \cG^{ijkl} D_i X_j (\dot g_{kl} - 2 D_k \dot M_l) = 0\ .
\end{multline}
Let us now go back to the steepest descent expression \eqref{tangexpr}
for $\delta S / \delta g_{ij}$ which we reproduce below for
convenience:
\begin{equation}\label{tangexpr1}
  \frac{\delta S}{\delta g_{ij}} = N^{-1}\cG^{ijkl} \big(g'_{kl} - 2 D_{(k} M'_{l)}\big) \frac{\ud S}{\ud s}\ .
\end{equation}
Using $\ud S / \ud s = \sqrt{-\epsilon V}$ from \eqref{ehjtangunit},
the definition of $\tau$ \eqref{dtau}, we get
\begin{equation}\label{tangexpr2}
  \frac{\delta S}{\delta g_{ij}} = \frac{1}{2N} 2\sqrt{-\epsilon V}\ \cG^{ijkl} \big(g'_{kl} - 2 D_{(k} M'_{l)}\big) = \frac{1}{2N}\cG^{ijkl} \big(\dot g_{kl} - 2 D_{(k} \dot M_{l)}\big)\ . 
\end{equation}
Substituting this back in the $\hvar N_i$ and $\hvar N$ equations of
motion \eqref{Neom}, we get
\begin{equation}\label{NNieom}
  - \cG_{ijkl} \frac{\delta S}{\delta g_{ij}} \frac{\delta S}{\delta g_{kl}} + \sqrt{g} (R - 2 \Lambda)   + N^{-1} 2 D_i X_j \frac{\delta S}{\delta g_{ij}} = 0\ ,\quad D_i \frac{\delta S}{\delta g_{ij}} = 0\ .
\end{equation}
Clearly, unless $X_i = 0$ (which is a solution of the equation
\eqref{Mieom}), the first equation does not match the original
Einstein-Hamilton-Jacobi equation \eqref{ehj}. Thus, we are led to the
choice $X_i = 0$, i.e., $N_i = \dot M_i$, for the $\hvar M_i$ equation
of motion. Here onwards, we substitute all occurrences of $\dot M_i$ by $N_i$:
\begin{equation}\label{NMrel}
  X_i = 0 \quad\Rightarrow\quad N_i = \dot M_i\ .
\end{equation}

\paragraph{Conjugate momentum and extrinsic curvature} In
Hamilton-Jacobi theory, the conjugate momentum $\pi^{ij}$ to the
configuration space variable $g_{ij}$ is defined as the derivative
$\delta S / \delta g_{ij}$ along the extremum path. Based on
\eqref{tangexpr2} and \eqref{NMrel}, we are led to the definition
\begin{equation}\label{pidef}
  \pi^{ij} = \frac{1}{2N} \cG^{ijkl} (\dot g_{kl} - 2 D_{(k} N_{l)})\ .
\end{equation}
The extremal path $g_{ij}(x,\tau)$ is treated as the evolution of the
3-manifold $\Sigma$ embedded as a hypersurface in four dimensional
spacetime, with $\tau$ parametrizing the different hypersurfaces along
the evolution. The following combination has a direct geometric
meaning as the \emph{extrinsic curvature} $K_{ij}$ of the embedded
$\Sigma$ in the $d+1$ dimensional space time:
\begin{equation}\label{Kdef}
  K_{ij} = \frac{1}{2N} \big(\dot g_{ij} - 2 D_{(i}  N_{j)}\big)\ ,
\end{equation}
and its trace is the \emph{mean curvature} $K$ of the embedded
$\Sigma$:
\begin{equation}\label{Ktrdef}
  K = g^{ij} K_{ij}\ .
\end{equation}
The extrinsic curvature is related to the conjugate momentum as
\begin{equation}
 \pi^{ij} = \cG^{ijkl} K_{kl} = \sqrt{g}(K^{ij} -  g^{ij} K)\ .
\end{equation}

\paragraph{The equations of motion} The variation of the action
$\hvar S$ with the above definitions in place is given by
\begin{align}\label{Svarfin}
  \hvar S &= \epsilon \int\ud\tau  \int_\Sigma \ud^{d}x\, \left( P^{ab} \hvar g_{ab} - \mc{H}_\perp \hvar N - \mc{H}^i \hvar N_i \right) + \epsilon \int\ud\tau \int_{\partial\Sigma} \ud^{d-1}\sigma\,r_i Q^i\ ,
\end{align}
with
\begin{align}\label{varcoeff}
  &\mc{H}_\perp = \cG_{ijkl} \pi^{ij} \pi^{kl} - \sqrt{g} (R-2\Lambda)\ ,\quad \mc{H}^i = -2 D_i \pi^{ij}\ ,\nonumber\\
  &P^{ab} =   - \frac{\partial\pi^{ab}}{\partial\tau} +   \sqrt{g} \big(-N (R^{ab} - \tfrac{1}{2} g^{ab} R + \Lambda g^{ab}) + D^a D^b N - g^{ab} D^c D_c N \big)  \nonumber\\
  &\quad\qquad - N \cG_{ijkl}{}^{,ab} \pi^{ij} \pi^{kl} -  D_i(\pi^{ib} N^a) -  D_j(\pi^{aj} N^b) +  D_p(\pi^{ab} N^p)\ ,
\end{align}
where $\cG_{ijkl}{}^{,ab} = \partial \cG_{ijkl} / \partial g_{ab}$,
the vector $r^i$ is the unit normal to the boundary $\partial\Sigma$,
and the boundary variations
\begin{align}\label{varboundary}
  Q^i = 2 N_j \hvar \pi^{ij} +  \cG^{ijkl} (ND_j \hvar g_{kl} - D_j N \hvar g_{kl}) + 2 \pi^{ij} N^p \hvar g_{pj} - N^i \pi^{ab} \hvar g_{ab}\ .
\end{align}
Setting $\hvar S = 0$ for arbitrary variations $\hvar g_{ij}$,
$\hvar N_i$, $\hvar N$ leads to the local equations of motion
$P^{ab} = \mc{H}_\perp = \mc{H}^i = 0$ when the boundary terms in
\eqref{Svarfin} are not present. When $\Sigma$ is closed, i.e.,
compact without boundary, the boundary terms \eqref{Svarfin} are
automatically absent. When $\Sigma$ has boundaries or asymptotic
regions, the boundary terms must be removed (1) by choosing
appropriate boundary conditions, and / or (2) by adding extra boundary
terms to the action $S$ whose variations cancel the terms in
\eqref{varboundary}, along the lines of \cite{Regge:1974zd,
  Henneaux:1984xu, Henneaux:1985tv, Brown:1992br,
  Hawking:1995fd}.\footnote{Indeed, the boundary terms
  \eqref{varboundary} are precisely the same terms (but with opposite
  sign) that
  \cite{Regge:1974zd,Henneaux:1984xu,Henneaux:1985tv,Brown:1992br,Hawking:1995fd}
  encounter in their analysis of the Hamiltonian of general relativity
  in the presence of boundaries on the Cauchy slice $\Sigma$. That the
  boundary terms here appear with opposite sign compared to
  \cite{Regge:1974zd,Henneaux:1984xu,Henneaux:1985tv,Brown:1992br,Hawking:1995fd}
  is consistent with the fact that here we consider the action whereas
  the above authors consider the Hamiltonian.}  Otherwise, when the
boundary variations are not zero, there are no solutions to the
variational principle since $\delta S = 0$ is never satisfied for
arbitrary bulk variations of fields.

Once the boundary terms have been handled, the variational principle
implies the following equations of motion:
\begin{align}
  & \cG_{ijkl} \pi^{ij} \pi^{kl} - \sqrt{g} (R-2\Lambda) = 0\ ,\quad -2 D_i \pi^{ij} = 0\ ,\label{conseq}\\
  & \frac{\ud g_{ij}}{\ud \tau} = 2N \cG_{ijkl} \pi^{kl} + 2 D_{(i}  N_{j)}\ ,\label{geq}\\
  & \frac{\partial\pi^{ab}}{\partial\tau} = -\sqrt{g} N (R^{ab} - \tfrac{1}{2} g^{ab} R + \Lambda g^{ab}) + \sqrt{g}(D^a D^b N - g^{ab} D^c D_c N)  \nonumber\\
  &\quad\qquad - N \cG_{ijkl}{}^{,ab} \pi^{ij} \pi^{kl} -  D_i(\pi^{ib} N^a) -  D_j(\pi^{aj} N^b) +  D_p(\pi^{ab} N^p)\ .\label{pieq}
\end{align}
We also include the equation for $\dot g_{ij}$ obtained by inverting
the definition of $\pi^{ij}$ \eqref{pidef}. The above equations are
precisely the Einstein equations written in Hamiltonian form.

\subsection{The ADM decomposition of the spacetime metric}

The above equations of motion describe a classical path in
configuration space of metrics modulo $d$-diffeomorphisms. The
parameter along the path is the `time' $\tau$ defined by
\begin{equation}\label{tausrel}
  \ud\tau = \frac{\ud s_\epsilon}{2\sqrt{-\epsilon V}}\ .
\end{equation}
If we assign a different spatial slice $\Sigma_\tau$ for each instant
$\tau$, with the metric on $\Sigma_\tau$ being $g_{ij}(x,\tau)$, then
the above path can be interpreted as a foliation of a $d+1$ dimensional
spacetime by the slices $\Sigma_\tau$ with the additional time
dimension being $\tau$. One can arrive at the notion of a metric on
this $d+1$ dimensional spacetime based on the equations
\eqref{conseq}-\eqref{pieq}:
\begin{equation}\label{ADMsplit}
\ud s_{d+1}^2 =  - N^2(x,\tau) \ud \tau^2 + g_{ij}(x,\tau) \big(\ud x^i + N^i(x,\tau) \ud \tau \big) \big(\ud x^j + N^j(x,\tau) \ud \tau \big) \ ,
\end{equation}
which is nothing but the ADM decomposition of a given metric on $d+1$
dimensional spacetime. Indeed, it is a standard exercise to plug in
the above formula \eqref{ADMsplit} into the $d+1$ dimensional Einstein
equations and obtain the equations of motion
\eqref{conseq}-\eqref{pieq}. Hence, our notion of time based on the
configuration space of $d$ dimensional metrics coincides with the
definition of time in the ADM decomposition for the spacetime
metric. We would like to reiterate that the formula for time above is
invariant under $d$-diffeomorphisms, i.e., independent of the choice
of coordinates on the spatial slice $\Sigma$.

Note that the formula \eqref{tausrel} is nothing but the Hamiltonian
constraint in disguise. To see this, start with the Hamiltonian
constraint
\begin{equation}
	\sqrt{g}(R-2\Lambda) = \cG^{ijkl} K_{ij} K_{kl} = (2N)^{-2} \cG^{ijkl} \big( \dot g_{ij} - 2 D_{i} \dot M_{j} \big) \big( \dot g_{kl} - 2 D_{k} \dot M_{l} \big) .
\end{equation}
Now we integrate this over all space:
\begin{equation}
	\epsilon\int \ud^{d}x\, N^{-1} \cG^{ijkl} \left( \dot g_{ij} - 2 D_{i} \dot M_{j} \right) \left( \dot g_{kl} - 2 D_{k} \dot M_{l} \right) = 4 \epsilon \int \ud^{d}x\, N \sqrt{g} (R-2\Lambda)\ ,
\end{equation}
Reparametrizing the path with an arbitrary parameter $\lambda$, the
left hand side of the above equation changes appropriately:
\begin{multline}
  \left(\frac{\ud\lambda}{\ud\tau}\right)^2 \epsilon \int_\Sigma \ud^{d}x\, N^{-1} \cG^{ijkl} \left(\frac{\ud g_{ij}}{\ud \lambda} - 2 D_{i} \frac{\ud M_{j}}{\ud\lambda}\right)\left(\frac{\ud g_{kl}}{\ud \lambda} - 2 D_{k} \frac{\ud M_{l}}{\ud\lambda}\right) \\  = 4  \epsilon \int \ud^{d}x\, N \sqrt{g} (R-2\Lambda)\ .
\end{multline}
Multiplying the above equation by $\ud\tau^2$ and using the definition
\eqref{orbmet} of $\ud s_\epsilon^2$ along the path, the above becomes the
relation $\ud s_\epsilon^2 = -4 \epsilon V\,\ud\tau^2$ sought above.\\

\begin{centering}
  {\small \emph{...And the end of all our exploring}\\
 \emph{Will be to arrive where we started}\\
 \emph{And know the place for the first time.}\\
\quad\quad {\small -- T. S. Eliot (The Four Quartets)}}\\
\end{centering}

\subsection{Matter degrees of freedom}

The above formula for time $\tau$ can be extended to include any
matter degrees of freedom that are minimally coupled to the
metric.\footnote{This discussion also applies when we have empirical
  sources that couple to the metric via their energy-momentum tensor.}
When matter degrees of freedom are present, the Hamilton-Jacobi
equations \eqref{ehjsmear} will include additional contributions from
the matter.  We illustrate the discussion for a scalar field $\phi(x)$
with potential $U(\phi)$. The Einstein-Hamilton-Jacobi equation
\eqref{ehjsmear} is modified to
\begin{equation}\label{ehjmatter}
  \int_\Sigma \ud^{d}x\, N \left(\cG_{ijkl} \frac{\delta S}{\delta g_{ij}} \frac{\delta S}{\delta g_{kl}} + \frac{1}{2\sqrt{g}}\frac{\delta S}{\delta \phi}\frac{\delta S}{\delta \phi} -  \sqrt{g} \big(R - 2\Lambda\big) + \frac{1}{2} \sqrt{g} g^{ij} \partial_i \phi \partial_j \phi+ \sqrt{g} U(\phi)\right) = 0\ ,
\end{equation}
whereas the $d$-diffeomorphism constraint is modified to
\begin{equation}
  -2 D_i \frac{\delta S}{\delta g_{ij}} + D^j \phi \frac{\delta S}{\delta \phi} = 0\ .
\end{equation}
The configuration space is now composed of the metric degrees of
freedom $g_{ij}(x)$ and the scalar degrees of freedom $\phi(x)$. The
metric on the configuration space is then the sum of the de Witt
metric and the scalar field metric that can be extracted from
\eqref{ehjmatter}:
\begin{multline}\label{mattermet}
\ud s^2 = \int_\Sigma \ud^{d}x\, N^{-1} \Big(\cG^{ijkl} (\delta g_{ij} - 2 D_{(i} \delta M_{j)}) (\delta g_{kl}- 2 D_{(k} \delta M_{l)}) \\ + 2 \sqrt{g} (\delta \phi - \delta M^iD_i\phi) (\delta \phi - \delta M^j D_j\phi)\Big)\ .
\end{multline}
Again, the distance measured with the above line element along a path
can be positive, negative or zero. Restricting ourselves to paths
along which $\ud s^2$ is non-zero, we define the positive line element
$\ud s_\epsilon^2 = |\ud s^2|$, with $\epsilon = \sign(\ud s^2)$ as
earlier. Following the same steps as earlier, the expression for the
Hamilton principal function on the extremal path comes out to be
\begin{equation}
  S = \int_{s_{\epsilon,0}}^{s_{\epsilon,1}} \ud s_\epsilon\left(\sqrt{-\epsilon V[g_{ij},N,\phi]} + \epsilon C[g_{ij}, M_i, N, N_i, \phi]\right)\ ,
\end{equation}
with
\begin{align}\label{VCdefmatter}
  V[g_{ij},N,\phi] &= -\int_\Sigma \ud^{d}x\, N\sqrt{g} \left( (R-2\Lambda) - \tfrac{1}{2} g^{ij} \partial_i \phi \partial_j\phi - U(\phi)\right)\ ,\\
  C[g_{ij},M_i,N,N_i,\phi] &= \int_\Sigma \ud^{d}x\,   N_{j} \Big(D_i\! \big(N^{-1} \cG^{ijkl}( g'_{kl} - 2 D_{k}  M'_{l})\big)\nonumber\\
                   &\qquad\qquad\qquad\qquad\qquad \qquad - N^{-1} \sqrt{g} (\phi'-M'^i D_i\phi) D^j \phi\Big)\ .
\end{align}
The equation for the extremal path is obtained by setting the path
variation $\hvar S$ to zero. Repeating the analysis Section
\ref{newtau}, the new time parameter $\tau$ is defined via
\begin{equation}
  \ud\tau = \frac{\ud s_\epsilon}{2\sqrt{-\epsilon V}}\ ,
\end{equation}
where $V$ given by \eqref{VCdefmatter}. As earlier, the $M_j$ equation
of motion is solved by $N_j = \dot M_j$ (recall that $\dot{}$ stands
for $\ud/\ud\tau$). The $N$ and $N^i$ equations of motion give the
Hamiltonian and momentum constraints with matter contributions:
\begin{equation}
\cG_{ijkl} \pi^{ij} \pi^{kl} + \frac{1}{2\sqrt{g}} \pi_\phi^2 - \sqrt{g} \big(R-2\Lambda - \tfrac{1}{2} \partial_i\phi \partial^i\phi - U(\phi) \big) = 0\ ,\quad -2 D_i\pi^{ij} + \pi_\phi D^j \phi = 0\ ,
\end{equation}
where the conjugate momenta are defined as
\begin{equation}
  \pi^{ij} = \frac{1}{2N} \cG^{ijkl} (\dot g_{kl} - 2D_{(k} N_{l)})\ ,\quad \pi_\phi = N^{-1}\sqrt{g}(\dot\phi - N^i D_i\phi)\ .
\end{equation}
The $g_{ij}$ and $\phi$ equations of motion are respectively,
\begin{align}
  \frac{\ud \pi^{ab}}{\ud \tau}
  &= -N \sqrt{g} (R^{ab} - \tfrac{1}{2} g^{ab} R + \Lambda g^{ab}) + \sqrt{g}(D^a D^b N - g^{ab} D^c D_c N)\nonumber\\
  &\quad - N \cG_{ijkl}{}^{,ab} \pi^{ij} \pi^{kl} - D_i(\pi^{ib} N^a) - D_j(\pi^{aj}  N^b) + D_p(\pi^{ab} N^p)\nonumber\\
  &\quad + \frac{1}{4\sqrt{g}} N g^{ab} \pi_\phi^2 + \frac{1}{2}N \Big(\partial^a\phi\partial^b\phi - g^{ab}\big(\tfrac{1}{2} g^{ij}\partial_i\phi\partial_j\phi + U(\phi)\big)\Big)\ ,\\
  \frac{\ud\pi_\phi}{\ud\tau} &= \partial_i (N \sqrt{g} g^{ij} \partial_j\phi) - \sqrt{g} N U'(\phi) + D_j(N^j \pi_\phi)\ .
\end{align}

\subsection{Euclidean signature spacetimes}\label{euclidsign}

The Einstein-Hamilton-Jacobi equation for Euclidean signature
spacetimes differs by a sign in the term quadratic in
$\delta S / \delta g_{ij}$, which can be thought of arising from the
standard Wick rotation of the canonical momentum
$\pi^{ij} \to \i \pi^{ij}$, since
$\frac{\delta S}{\delta g_{ij}} = \pi^{ij}$ on classical paths (see
\eqref{pidef}). Thus, the Einstein-Hamilton-Jacobi equation is
\begin{equation}
  \wt{\epsilon}\, \cG_{ijkl} \frac{\delta S}{\delta g_{ij}} \frac{\delta S}{\delta g_{kl}}  - \sqrt{g}(R - 2\Lambda) = 0\ ,\quad \wt{\epsilon} = \left\{\begin{array}{cc} +1 & \text{for Lorentzian spacetimes} \\ -1 & \text{for Euclidean spacetimes}\end{array}\right.
\end{equation}
Note that the signature of the configuration space metric $\cG^{ijkl}$
is still indefinite since the configuration space of metrics on the
three dimensional $\Sigma$ is the same for both Euclidean and
Lorentzian evolution. The norm-squared of the tangent vector to a
classical path can still be zero, positive or negative, and is
characterized by the sign $\epsilon$. The rest of the calculation
proceeds as before: for instance, the analog of \eqref{ehjtangunit} is
\begin{equation}\label{EHJsimpleeuclid}
  \epsilon   \wt{\epsilon}\left(\frac{\ud S}{\ud s_\epsilon}\right)^2 = - V[g_{ij}, N] = \int_\Sigma \ud^{d}x \sqrt{g} N (R - 2\Lambda) \ ,
\end{equation}
which gives $\epsilon \wt\epsilon = \sgn(-V)$. Thus, every appearance
of the sign $\epsilon$ is replaced by $\wt\epsilon \epsilon$.

\section{Illustrating the formula for `time' $\tau$ for de Sitter
  spacetime} \label{examples}

In this section, we look at a few simple examples to illustrate the
notion of time $\tau$ given by the following universal formula in
terms of the configuration space variables:
\begin{equation}\label{taudef1}
  \ud \tau = \frac{\ud s_\epsilon}{2\sqrt{-\epsilon V}}\ .
\end{equation}

\subsection{Recap}

We recall the definition of the various terms in the formula
above. The line element $\ud s_\epsilon$ along a path in configuration
space of metrics $\mc{M}_\Sigma$ is given by \eqref{orbmet}:
\begin{align}\label{orbmet1}
  \ud s_\epsilon^2 &=  \epsilon\, \ud s^2\ ,\quad\text{with}\quad \ud s^2 = \int_\Sigma \ud^{d}x\, N^{-1} \cG^{ijkl} \left(\delta g_{ij} - 2 D_{i} \delta M_{j}\right)\left(\delta g_{kl} - 2 D_{k}\delta M_{l}\right)\ ,
\end{align}
where $\delta g_{ij} - 2 D_{i} \delta M_{j}$ is along the tangent to
the classical path, $\epsilon$ is the sign of $\ud s^2$ along the
classical path which we restrict to be non-zero, $N$ is the positive,
nowhere zero lapse function, and $\cG^{ijkl}$ is the de Witt metric
\eqref{deWitt}. Recall from equation \eqref{ehjtangunit} that we also have

\begin{equation}\label{potdef}
  \epsilon = \sgn(-V)\ ,\quad \text{with}\quad V = -\int_\Sigma \ud^{d}x\,N \sqrt{g} (R-2\Lambda)\ .
\end{equation}
It will also be useful to recall the decomposition
\eqref{deWittdecomp} of the de Witt metric in terms of the conformal
mode $\Omega = g^{1/4}$ of $g_{ij}$ and
$\tl{g}_{ij} = g^{-1/3} g_{ij}$ which has $\det \tl{g}_{ij} = 1$:
\begin{equation}\label{deWittdecomp2}
  \cG^{ijkl}(x) \delta g_{ij}(x) \delta g_{kl}(x) = -\frac{32}{3} \delta \Omega(x)^2 + \Omega^2(x)\, \tl{g}^{ij} \delta \tl{g}_{jk}(x)\, \tl{g}^{kl} \delta \tl{g}_{li}(x)\ .
\end{equation}

\subsection{Global de Sitter spacetime}\label{dSsec}

Four dimensional de Sitter spacetime is a maximally symmetric space
with positive cosmological constant $\Lambda > 0$, and symmetry group
$\text{SO}(4,1)$. This can be seen from the definition of de Sitter
space as a hyperboloid in $\mathbf{R}^{4,1}$:
\begin{equation}\label{dsembed}
  -X_0^2 + X_1^2 + X_2^2 + X_3^2 + X_4^2 = \ell^2\ ,
\end{equation}
which satisfies Einstein's equations $R_{\mu\nu} = \Lambda g_{\mu\nu}$
with $\Lambda = 3 / \ell^2$. This spacetime can be viewed as the time
development of the metric on a three dimensional spatial slice
$\Sigma$ in many different ways. To illustrate the notion of time
provided by \eqref{taudef1} in the simplest possible situation, we can
restrict the 3-metrics to be invariant under as big a subgroup of
$\text{SO}(4,1)$ as possible. Restricting the metric this way cuts
down on the number of free components of the metric drastically. For
instance, choosing the spatial metric to be invariant under the
$\text{SO}(4)$ subgroup of $\text{SO}(4,1)$, the spatial slice must be
a 3-sphere $\mathbf{S}^3$ of some radius, so that there is only one
parameter viz., the radius, which can change over the time
development.\footnote{The group $\text{SO}(4)$ has the maximum number
  of symmetries (6) for a metric in three dimensions. There are two
  other maximally symmetric metrics in three dimensions and we look at
  these in Section \ref{dsothermax}.}
The metric $g_{ij}$ on $\Sigma$ is then the round metric on
$\mathbf{S}^3$ of radius $a$:
\begin{equation}
  g_{ij} \ud x^i \ud x^j = a^2 \left(   \ud\chi^2 + \sin^2\chi (\ud\theta^2 + \sin^2\theta\ud\varphi^2)\right)\ .
\end{equation}
The above is a one-parameter family of metrics, parametrized by the
radius $a$. Thus, the only dynamical variable is the constant
conformal mode
$\Omega = g^{1/4} = a^{3/2} \sin\chi \sqrt{\sin \theta}$ and the
smeared de Witt metric is 
\begin{equation}
  \ud s^2 = -48\pi^2 N^{-1} a\, \ud a^2\ .
\end{equation}
where we have taken the lapse to be independent of $\chi$, $\theta$
and $\varphi$ as well, due to the assumption of maximal symmetry on
$\Sigma$. Thus, the sign $\epsilon = \sign (\ud s^2) = -1$ so that
\begin{equation}
  \ud s_\epsilon^2 = 48\pi^2  N^{-1}\, a \ud a^2\ ,
\end{equation}
The potential $V$ is also simple to compute since the
curvature $R$ of the metric $g_{ij}$ is constant on $\Sigma$ due to
maximal symmetry: $R = 6 a^{-2}$. We then have
\begin{equation}\label{VdeSitter}
  -V = \int_\Sigma N\sqrt{g}(R-2\Lambda) = 12\pi^2 N a\left(1 - \frac{a^2}{\ell^2}\right)\ ,
\end{equation}
where we have used $\ell^2 = 3 / \Lambda$. The definition of $\tau$ is
then
\begin{equation}
  \ud \tau = \frac{\ud s_\epsilon}{2\sqrt{-\epsilon V}}  =  N^{-1} \frac{ \ud a}{\sqrt{\frac{a^2}{\ell^2}-1}}\ ,
\end{equation}
which gives
\begin{equation}
\int_{0}^\tau N \ud \tau = \int_{\ell}^a \frac{\ud a}{\sqrt{\frac{a^2}{\ell^2}-1}} = \ell \cosh^{-1} \frac{a}{\ell} = \ell \log\left(\frac{a}{\ell} + \sqrt{\frac{a^2}{\ell^2}-1}\right)\ ,
\end{equation}
where we have chosen the zero of time $\tau$ to coincide with
$a = \ell$. Suppose we choose $N = 1$, which is the same as working
with proper time $\int N \ud\tau$, we get
\begin{equation}
  \tau = \ell  \log\left(\frac{a}{\ell} + \sqrt{\frac{a^2}{\ell^2}-1}\right)\ .
\end{equation}
For $a \gg \ell$, we see that $\tau \sim \ell \log (2a/\ell)$, so that
the conformal mode gives a notion of time. More concretely, since the
conformal mode is independent of spatial coordinates, one can recast
the above relation in terms of the volume $\mc{V} = 2\pi^2 a^3$ of the
spatial slice:
\begin{equation}
  \tau \sim \frac{1}{3} \log \mc{V}\ .
\end{equation}
The above is the usual notion of time in de Sitter spacetime in terms
of the spatial volume\footnote{See \cite{Chakraborty:2023yed} for an
  application of this idea in understanding solutions of the
  Wheeler-de Witt equation in a closed universe with positive
  cosmological constant.}. In fact, we can express $a$ in terms of
$\tau$ as $a = \ell \cosh (\tau / \ell)$. Remembering that $N = 1$, we
indeed get the $3+1$ de Sitter metric:
\begin{equation}\label{dsmetric}
  -\ud \tau^2 + \ell^2 \cosh^2 \left(\frac{\tau}{\ell}\right) \big(\ud \chi^2 + \sin^2\chi(\ud\theta^2 + \sin^2\theta \ud\varphi^2)\big)\ .
\end{equation}
The $\mathbf{S}^3$ slicing of de Sitter space can be obtained by a
simple parametrization of the embedding coordinates $X_{0,\ldots,4}$
in \eqref{dsembed}:
\begin{equation}
  X_0 = \ell \sinh \frac{\tau}{\ell}\ ,\quad \sqrt{X_1^2 + X_2^2 + X_3^2 + X_4^2} = \ell \cosh \frac{\tau}{\ell}\ ,
\end{equation}
with the $X_{i=1,\ldots,4}$ constrained to be on an $\mathbf{S}^3$
with radius $\ell \cosh (\tau / \ell)$ and coordinates $\chi$,
$\theta$, $\varphi$ \eqref{dsmetric}.

\subsection{Other maximally symmetric slicings}\label{dsothermax}
  
There are in fact three possibilities for maximally symmetric metrics
in three euclidean dimensions corresponding to the isometry groups
$\text{SO}(4)$, $\text{SO}(3,1)$ and $\text{ISO}(3)$ (the three
dimensional Poincare group), which are all subgroups of
$\text{SO}(4,1)$. The three manifolds corresponding to these are the
round 3-sphere $\mathbf{S}^3$, the hyperbolic space $\mathbf{H}^3$ and
euclidean space $\mathbf{R}^3$ respectively. We have already looked at
the $\mathbf{S}^3$ slicing above. Here, we look at the remaining two
possibilities. The standard metric on these spaces is
\begin{align}\label{maxsymmet}
  \mathbf{H}^3:&\quad \ud\psi^2 + \sinh^2\psi (\ud\theta^2 + \sin^2\theta\ud\varphi^2)\ ,\nonumber\\
  \mathbf{R}^3:&\quad \ud\rho^2 + \rho^2 (\ud\theta^2 + \sin^2\theta\ud\varphi^2)\ .
\end{align}
Here, $\rho$ and $\psi$ are radial coordinates, and $\theta$,
$\varphi$ are standard angular coordinates with ranges
$0 \leq \theta \leq \pi$, $0 \leq \varphi < 2\pi$. These metrics have
infinite volumes and have to be renormalized: we denote the
renormalized volume by $\vol_0$. As we shall see, this renormalized
volume drops out of our formula for the time $\tau$, and hence we do
not need to know the details of the renormalization\footnote{For a
  detailed treatment of the divergences and the required counterterms,
  see, for instance,
  \cite{Balasubramanian:1999re,Emparan:1999pm,Lau:1999dp,Mann:1999pc,Kraus:1999di}.}.

The metric $g_{ij}$ on $\Sigma$ is then $g_{ij} = a^2 \hat{g}_{ij}$
where $\hat{g}_{ij}$ stands for either of the metrics in
\eqref{maxsymmet}. The conformal mode is
$\Omega = a^{3/2} \hat{g}^{1/4}$ where $\hat{g}$ is the determinant of
the metric \eqref{maxsymmet}. As earlier, the constant conformal mode
is the only degree of freedom active along the path so that
$\epsilon = -1$. We also choose the lapse $N = 1$ for simplicity. We
then get $\ud s_\epsilon^2 = 24 \vol_0\, a \ud a^2$.
The scalar curvature is $R = 6k / a^2$ where $k = -1$ for
$\mathbf{H}^3$ and $k = 0$ for $\mathbf{R}^3$, so that
$V = -6 a \vol_0 ( k - \frac{a^2}{\ell^2})$.
The time $\tau$ is then defined by
\begin{equation}
  \ud\tau = \frac{\ud s_\epsilon}{2\sqrt{-\epsilon V}} = \frac{\ud a}{\sqrt{-k + \frac{a^2}{\ell^2}}}\ ,
\end{equation}
where recall that $\ell^2 = 3 / \Lambda$. Note that the renormalized
volume $\vol_0$ has cancelled between the numerator and
denominator. Integrating the above equation, we get the following
formula for $\tau$ in terms of the constant conformal mode $a$ for
$k = -1$ and $k = 0$:
\begin{align}
  \text{For $\mathbf{H}^3$ slicing}:&\quad  \frac{\tau}{\ell} =  \log \left(\sqrt{\frac{a^2}{\ell^2} + 1} + \frac{ a}{\ell}\right)\ ,\nonumber\\
  \text{For $\mathbf{R}^3$ slicing}:&\quad  \frac{\tau}{\ell} = \log \frac{a}{\ell}\ .
\end{align}
Inverting the above relation for $a$ in terms of $\tau$, we get the
following metric for de Sitter space with $\mathbf{H}^3$ slicing and
flat slicing respectively:
\begin{align}
  \text{For $\mathbf{H}^3$ slicing}:&\quad  -\ud \tau^2 + \ell^2 \sinh^2 \left(\frac{\tau}{\ell}\right) \left(\ud\psi^2 + \sinh^2 \psi (\ud\theta^2 + \sin^2\theta\ud\varphi^2)\right)\ ,\nonumber\\
  \text{For $\mathbf{R}^3$ slicing}:&\quad  -\ud \tau^2 +  \ell^2 \e^{2\tau / \ell}\left(\ud \rho^2 + \rho^2 (\ud\theta^2 + \sin^2\theta\ud\varphi^2)\right)\ .
\end{align}
The last metric is conformally flat, as can be seen from the
coordinate transformation $\eta = \e^{-\tau / \ell}$:
\begin{align}
  &-\ud \tau^2 +  \ell^2 \e^{2\tau / \ell}\left(\ud \rho^2 + \rho^2 (\ud\theta^2 + \sin^2\theta\ud\varphi^2)\right) =\frac{\ell^2}{\eta^2}\left(-\ud \eta^2 + \ud \rho^2 + \rho^2 (\ud\theta^2 + \sin^2\theta\ud\varphi^2)\right)\ .
\end{align}
The above non-compact slicings cover only a part of de Sitter space,
as can be seen by expressing the embedding coordinates \eqref{dsembed}
in terms of the above coordinates. See, for instance, the monograph
\cite{Griffiths:2009dfa} for the detailed coordinate transformations.

\subsection{Euclidean signature de Sitter spacetime}

Here, we look at Euclidean signature evolution of the $\mathbf{S}^3$
slice with positive cosmological constant $\Lambda$.  Recall from
Section \ref{euclidsign} that the Einstein-Hamilton-Jacobi equation
that incorporates both Euclidean and Lorentzian signature spacetimes
is
\begin{equation}
  \wt{\epsilon}\, \cG_{ijkl} \frac{\delta S}{\delta g_{ij}} \frac{\delta S}{\delta g_{kl}}  - \sqrt{g}(R - 2\Lambda) = 0\ ,\quad \wt{\epsilon} = \left\{\begin{array}{cc} +1 & \text{for Lorentzian spacetimes} \\ -1 & \text{for Euclidean spacetimes}\end{array}\right.
\end{equation}
We then have $\epsilon \wt\epsilon = \sgn(-V)$. For Euclidean
signature $\wt\epsilon = -1$, and for tangents only along the
conformal mode we have $\epsilon = -1$, so that $\sgn(-V) = +1$. This
is satisfied by $V$ in \eqref{VdeSitter} when $a^2 \leq \ell^2$. With
the choice $N=1$, the equation for $\tau$ becomes
\begin{equation}
  \ud\tau = \frac{\ud s_\epsilon}{2\sqrt{-\epsilon \tl\epsilon V}} = \frac{\ud a}{\sqrt{1 - \frac{a^2}{\ell^2}}}\quad\Rightarrow\quad a = \ell \sin\frac{\tau}{\ell}\ .
\end{equation}
The range of $\tau$ is from $0$ to $\pi\ell$ which corresponds to $a$
starting from $0$ at $\tau = 0$, reaching the maximum $\ell$ at
$\tau = \pi \ell / 2$ and ending at $0$ at $\tau = \pi\ell$. Defining
$\psi = \tau / \ell$, we get the four dimensional euclidean spacetime
metric
\begin{equation}
\ell^2 \Big(  \ud\psi^2 + \sin^2 \psi \big(\ud\chi^2 + \sin^2\chi (\ud \theta^2 + \sin^2\theta \ud\varphi^2)\big)\Big)\ ,
\end{equation}
which is nothing but the round metric on $\mathbf{S}^4$ with radius
$\ell$ (euclidean $\mathbf{dS}_4$).

\section{The time $\tau$ in asymptotically Anti de Sitter
  spacetimes}\label{adsexample}

It is an important result in the case of an AAdS spacetime that, with
some reasonable assumptions, it can always be foliated by spatial
slices which have zero mean curvature, i.e.,
$K \equiv g^{ij} K_{ij} = 0$
\cite{Wittentalk,Witten:2022xxp,Chrusciel:2022cjz}.\footnote{We thank
  E. Witten for pointing us to
  \cite{Witten:2022xxp}.}\textsuperscript{,}\footnote{For some more
  details about the spacetimes where maximal slicing is possible, see
  \cite{Witt:1986ng,Bartnik:1990um,Witt:2009za} and references
  therein. We thank G. Horowitz for pointing us to this set of
  references.} By the definition of the extrinsic curvature $K_{ij}$
\eqref{Kdef}, this corresponds to the tangent vector to the extremal
path being traceless. The smeared de Witt metric \eqref{deWittsmear}
is always positive definite on such traceless tangent vectors, in
which case the sign $\epsilon$ is \emph{always} $+1$. Thus, our
analysis in Section \ref{GRsec} is simpler in this case since we can
always ensure that $\epsilon = +1$ for classical paths which describe
AAdS spacetimes.

In forthcoming work, we address the issue of time in the important
case of black holes in AAdS spacetimes and its interpretation in the
dual conformal field theories.

\acknowledgments

We would like to thank Gary Horowitz, Alok Laddha, Gautam Mandal,
Kyriakos Papadodimas, R. Loganayagam, Suvrat Raju and Ashoke Sen for
discussions. This work was supported by DAE, Government of India,
under project no. RTI4001. S. R. W. would like to thank the Infosys
Foundation Homi Bhabha Chair at ICTS-TIFR for its support, the Theory
Division of CERN, Geneva where part of this work was done, and KITP,
Santa Barbara, for the stimulating program ``What is String Theory?
Weaving Perspectives Together'', supported by the NSF grant
PHY-2309135, where some aspects of this work were discussed with the
program participants.




\bibliographystyle{JHEP}
\bibliography{refs}

\end{document}